\documentclass[aip,jcp,reprint,noshowkeys,superscriptaddress]{revtex4-1}
\usepackage{graphicx,dcolumn,bm,xcolor,multirow,amscd,amsmath,amssymb,amsfonts,physics,longtable,wrapfig,bbold,xspace,microtype,siunitx}
\usepackage[version=4]{mhchem}

\usepackage[utf8]{inputenc}
\usepackage[T1]{fontenc}

\usepackage{hyperref}
\hypersetup{
    colorlinks,
    linkcolor={red!50!black},
    citecolor={red!70!black},
    urlcolor={red!80!black}
}

\usepackage{listings}
\definecolor{codegreen}{rgb}{0.58,0.4,0.2}
\definecolor{codegray}{rgb}{0.5,0.5,0.5}
\definecolor{codepurple}{rgb}{0.25,0.35,0.55}
\definecolor{codeblue}{rgb}{0.30,0.60,0.8}
\definecolor{backcolour}{rgb}{0.98,0.98,0.98}
\definecolor{mygray}{rgb}{0.5,0.5,0.5}

\definecolor{sqred}{rgb}{0.85,0.1,0.1}
\definecolor{sqgreen}{rgb}{0.25,0.65,0.15}
\definecolor{sqorange}{rgb}{0.90,0.50,0.15}
\definecolor{sqblue}{rgb}{0.10,0.3,0.60}

\lstdefinestyle{mystyle}{
    backgroundcolor=\color{backcolour},
    commentstyle=\color{codegreen},
    keywordstyle=\color{codeblue},
    numberstyle=\tiny\color{codegray},
    stringstyle=\color{codepurple},
    basicstyle=\ttfamily\footnotesize,
    breakatwhitespace=false,
    breaklines=true,
    captionpos=b,
    keepspaces=true,
    numbers=left,
    numbersep=5pt,
    numberstyle=\ttfamily\tiny\color{mygray},
    showspaces=false,
    showstringspaces=false,
    showtabs=false,
    tabsize=2
  }

  \newcolumntype{d}{D{.}{.}{-1}}

\usepackage[normalem]{ulem}
\newcommand{\titou}[1]{\textcolor{black}{#1}}

\newcommand{\SupMat}{\textcolor{blue}{supplementary material}}

\newcommand{\fnm}{\footnotemark}
\newcommand{\fnt}{\footnotetext}
\newcommand{\tabc}[1]{\multicolumn{1}{c}{#1}}

\newcommand{\T}[1]{#1^{\intercal}}
\newcommand{\ii}{\textrm{i}}

\newcommand{\br}{\boldsymbol{r}}
\newcommand{\bx}{\boldsymbol{x}}

\newcommand{\HF}{\text{HF}}

\newcommand{\eh}{\text{eh}}
\newcommand{\pp}{\text{pp}}
\newcommand{\ee}{\text{ee}}
\newcommand{\hh}{\text{hh}}

\newcommand{\e}{\epsilon}
\newcommand{\Om}{\Omega}

\newcommand{\MO}{\varphi}

\newcommand{\bO}{\boldsymbol{0}}
\newcommand{\bId}{\boldsymbol{1}}

\newcommand{\bOm}{\boldsymbol{\Omega}}

\newcommand{\bA}{\boldsymbol{A}}
\newcommand{\bB}{\boldsymbol{B}}
\newcommand{\bC}{\boldsymbol{C}}

\newcommand{\bX}{\boldsymbol{X}}
\newcommand{\bY}{\boldsymbol{Y}}
\newcommand{\bU}{\boldsymbol{U}}

\newcommand{\ov}{\overline{v}}
\newcommand{\oV}{\overline{V}}
\newcommand{\oOm}{\overline{\Om}}
\newcommand{\oA}{\overline{A}}
\newcommand{\oB}{\overline{B}}
\newcommand{\oX}{\overline{X}}
\newcommand{\oY}{\overline{Y}}
\newcommand{\obOm}{\overline{\bOm}}
\newcommand{\obA}{\overline{\bA}}
\newcommand{\obB}{\overline{\bB}}
\newcommand{\obX}{\overline{\bX}}
\newcommand{\obY}{\overline{\bY}}

\newcommand{\cre}[1]{a_{#1}^\dagger}
\newcommand{\ani}[1]{a_{#1}}

\newcommand{\up}{\uparrow}
\newcommand{\dw}{\downarrow}
\newcommand{\upup}{\uparrow\uparrow}
\newcommand{\updw}{\uparrow\downarrow}
\newcommand{\dwup}{\downarrow\uparrow}
\newcommand{\dwdw}{\downarrow\downarrow}

\DeclareSIUnit{\angstrom}{\textup{\AA}}

\newcommand{\lcpq}{Laboratoire de Chimie et Physique Quantiques, Universit\'e de Toulouse, CNRS, UPS, France}
\newcommand{\lpt}{Laboratoire de Physique Th\'eorique, Universit\'e de Toulouse, CNRS, UPS, France}
\newcommand{\etsf}{European Theoretical Spectroscopy Facility (ETSF)}

\begin{document}	

\title{The three channels of many-body perturbation theory: $GW$, particle-particle, and electron-hole $T$-matrix self-energies}

\author{Roberto Orlando}
    \affiliation{\lcpq}
    \affiliation{\lpt}
    \affiliation{\etsf}
\author{Pina Romaniello}
\email{pina.romaniello@irsamc.ups-tlse.fr}
    \affiliation{\lpt}
    \affiliation{\etsf}
\author{Pierre-Fran\c{c}ois Loos}
\email{loos@irsamc.ups-tlse.fr}
    \affiliation{\lcpq}

\begin{abstract}
We derive the explicit expression of the three self-energies that one encounters in many-body perturbation theory: the well-known $GW$ self-energy, as well as the particle-particle and electron-hole $T$-matrix self-energies.
Each of these can be easily computed via the eigenvalues and eigenvectors of a different random-phase approximation (RPA) linear eigenvalue problem that completely defines their corresponding response function.
For illustrative and comparative purposes, we report the principal ionization potentials of a set of small molecules computed at each level of theory.
\titou{The performance of these schemes on strongly correlated systems (\ce{B2} and \ce{C2}) is also discussed.}
\end{abstract}

\maketitle
\section{The standard form of Hedin's equations}
\label{sec:hedin}

The quasiparticle picture is a central concept in quantum many-body physics and chemistry as it provides a means to understand the behavior of electrons within a material or a molecule. \cite{CsanakBook,FetterBook,MattuckBook,MartinBook} It emerges as an effective mapping from the complex many-body system to a simplified effective one-body system. Within the quasiparticle framework, the effects of collective excitations are incorporated by adding a dynamical correction to an effective one-body operator obtained from a simpler system, such as the non-interacting system. This correction, which contains Hartree (H), exchange (x), and correlation (c) effects, is known as the self-energy and is denoted as $\Sigma$. The famous Hedin equations, a self-consistent set of five integrodifferential equations, provide a route to calculate this self-energy. \cite{Hedin_1965} Their conventional form is
\begin{subequations}
\begin{align}
	\label{eq:Gamma}
	\begin{split}
	\Gamma(123) & = \delta(12) \delta(13) 
	\\
	& + \Xi_\text{xc}(12;45) G(46) G(75) \Gamma(673)
	\end{split}
	\\
	\label{eq:P}
	P(12) & = - \ii G(13) G(41) \Gamma(342)
	\\
	\label{eq:W_2p}
	W(12) & = v(12) + v(13) P(34) W(42)
	\\
	\label{eq:Sigma_xc}
	\Sigma_\text{xc}(12) & = \ii G(14) W(1^+3) \Gamma(423)
	\\
	\label{eq:G}
	G(12) & = G_\text{H}(12) + G_\text{H}(13) \Sigma_\text{xc}(34) G(42)
\end{align}
\end{subequations}
where $P$ is the \textit{irreducible} polarizability, $W$ and $v$ are the dynamically screened and bare Coulomb interactions, $\Gamma$ is the \textit{irreducible} three-point vertex which is completely defined by the four-point exchange-correlation kernel
\begin{equation}
	\Xi_\text{xc}(12;1'2') = \fdv{\Sigma_\text{xc}(11')}{G(2'2)}
\end{equation}
In these equations, integrals over repeated indices are assumed, and, for instance, $1 = (\br_1,\sigma_1,t_1)$ is a space-spin-time variable and $1^+ = (\br_1,\sigma_1,t_1+\delta)$ with $\delta \to 0^+$. $G$ and $G_\text{H}$ are the fully-interacting and Hartree Green's functions, respectively, and are linked by a Dyson equation, Eq.~\eqref{eq:G}. The exchange-correlation part of the self-energy is
\begin{equation}
	\Sigma_\text{xc} = \Sigma_\text{x} + \Sigma_\text{c} = \Sigma - \Sigma_\text{H}
\end{equation}
where the Hartree and exchange components are respectively given by
\begin{subequations}
\begin{align}
	\label{eq:SigH}
	\Sigma_\text{H}(12) & = - \ii \delta(12) v(1^+3) G(33^+)
	\\
	\label{eq:SigX}
	\Sigma_\text{x}(12) & = + \ii v(1^+2) G(12)
\end{align}
\end{subequations}
with $\delta$ the Dirac delta function.

\begin{figure}
	\includegraphics[width=0.45\linewidth]{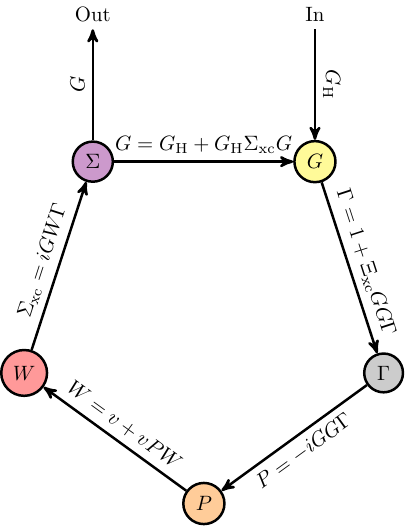}
	\hspace{0.07\linewidth}
	\includegraphics[width=0.45\linewidth]{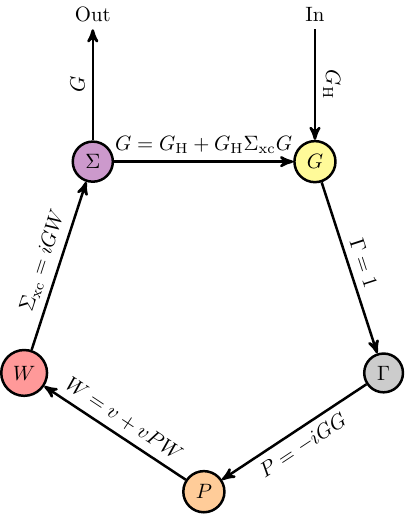}
	\caption{Left: Hedin's pentagon yielding the exact Green's function $G$. Right: Hedin's pentagon yielding the $GW$ approximation by setting $\Gamma = 1$ in the expression of $P$ and $\Sigma_\text{xc}$.}
	\label{fig:pentagon}
\end{figure}

As shown schematically in Fig.~\ref{fig:pentagon}, one can easily obtain the $GW$ \textit{form} of the self-energy from Hedin's equations by neglecting the vertex corrections, i.e., by setting $\Gamma(123) = \delta(12) \delta(13)$ in Eq.~\eqref{eq:Gamma}, yielding $\Sigma_\text{xc}(12) = \ii G(12) W(12)$. \cite{Aryasetiawan_1998,Onida_2002,Reining_2017,Golze_2019} The $GW$ \textit{approximation} is obtained by additionally setting $\Gamma(123) = \delta(12) \delta(13)$ in the expression of the irreducible polarizability given in Eq.~\eqref{eq:P},which reads $P^\text\eh(12) = - \ii G(12)G(21)$. Diagrammatically, the $GW$ equations correspond to a resummation of the direct ring (or bubble) diagrams \cite{Gell-Mann_1957,MattuckBook} and its central quantity is the two-point dynamically screened Coulomb interaction $W(12) = v(12) + v(13) P^\text\eh(34) W(42)$.

Other types of diagrams can be resummed, such as ladder diagrams. \cite{MartinBook,MattuckBook} This alternative resummation defines the $T$-matrix approximation that has the effective four-point interaction $T(12;1'2')$ as a key object. \cite{Baym_1961,Baym_1962,Danielewicz_1984a,Danielewicz_1984b} The two types of ladder diagrams, electron-hole (eh) and particle-particle (pp), produce two different channels for the $T$-matrix that one can write down in terms of Dyson equations with a random-phase approximation (RPA) \cite{Bohm_1951,Pines_1952,Bohm_1953,Nozieres_1958} polarizability as a kernel. It is however not convenient to derive the $T$-matrix approximation from the conventional form of Hedin's equations.

\section{An alternative form of Hedin's equations}
\label{sec:hedin_bis}

Following Romaniello and coworkers \cite{Romaniello_2012} (see also Refs.~\onlinecite{Starke_2012,Maggio_2017b} for an alternative derivation), one can recast Hedin's equations in a more convenient way by considering the Dyson equation that links the non-interacting two-body correlation function
\begin{equation}
	L_0(12;1'2') = G(12') G(21')
\end{equation}
to the full two-body correlation function  
\begin{equation}
\label{eq:BSE}
\begin{split}
	L(12;1'2') 
	& = L_0(12;1'2') 
	\\ 
	& + L_0(14;1'3) \Xi(35;46) L(62;52')
\end{split}
\end{equation}
with
\begin{equation}
\label{eq:kernel}
\begin{split}
	\Xi(12;1'2') 
	& = \fdv{\Sigma(11')}{G(2'2)}
	= \Xi_\text{H}(12;1'2') + \Xi_\text{xc}(12;1'2')
	\\
	& = - \ii V(12;1'2') + \Xi_\text{xc}(12;1'2')
\end{split}
\end{equation}
where it is convenient to introduce, at this stage, the four-point version of the bare and dynamically screened Coulomb interactions
\begin{subequations}
\begin{align}
	V(12;1'2') & = \delta(11') \delta(22') v(12)
	\\
	W(12;1'2') & = \delta(12') \delta(1'2) W(12)
\end{align}
\end{subequations}

\begin{figure*}
	\hspace{0.1\linewidth}
	\includegraphics[width=0.2\linewidth]{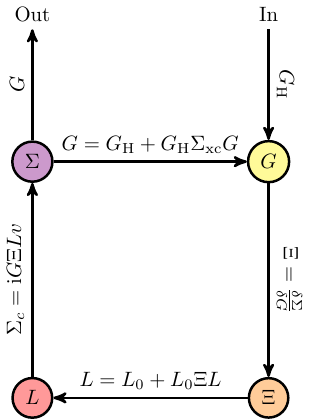}
	\hspace{0.08\linewidth}
	\includegraphics[width=0.2\linewidth]{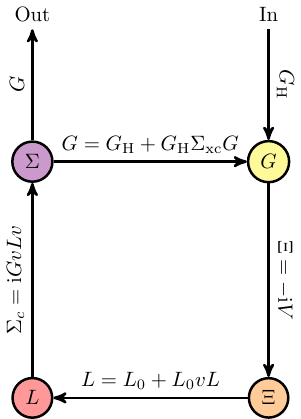}	
	\hspace{0.08\linewidth}
	\includegraphics[width=0.2\linewidth]{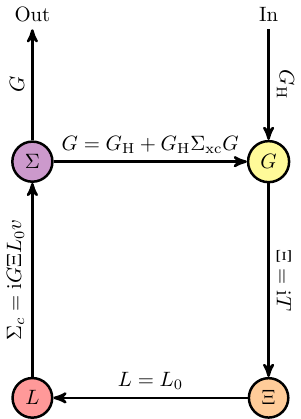}	
	\hspace{0.1\linewidth}
	\caption{Left: Hedin's ``square'' yielding the exact Green's function $G$. Center: Hedin's square yielding the $GW$ approximation by setting $\Xi = - \ii V$. Right: Hedin's square yielding the $T$-matrix approximation by setting $L = L_0$ and $\Xi = \ii T$.}
	\label{fig:square}
\end{figure*}

Equation \eqref{eq:BSE} is the Bethe-Salpeter equation of the two-body correlation function which completely defines the dynamically screened interaction via 
\begin{equation}
\begin{split}
	\label{eq:W_BSE}
	W(12;1'2') 
	& = V(12;2'1') 
	\\
	& - \ii V(13;2'3') L(34;3'4') V(42;4'1')
\end{split}
\end{equation}
The latter equation is the four-point extension of the two-point expression $W(12)=v(12)+v(13)\chi(34)v(42)$, with $\chi(12)=-\ii L(12;1^+2^+)$ the response function, which can be obtained from Eq.~\eqref{eq:W_2p} through the link $-\ii L=(1-vP)^{-1}P$.
Together with Eqs ~\eqref{eq:BSE} and \eqref{eq:kernel}, we obtain a more compact form of Hedin's equations (see Fig.~\ref{fig:square}):
\begin{subequations}
\begin{align}
	\Sigma_{c}(12) & = \ii G(13) \Xi(35;26) L(64;54) v(14)
	\label{eq:Sig}
	\\
	G(12) & = G_\text{Hx}(12) + G_\text{Hx}(13) \Sigma_\text{c}(34) G(42)
\end{align}
\end{subequations}
A different derivation of the standard and alternative forms of Hedin's equation is presented in Appendix \ref{app:hedin_alt} based on the equation-of-motion formalism.

\section{Dyson equations}
\label{sec:dyson}

We are now in a position to explain how to obtain the $GW$ and $T$-matrix expressions of the self-energy building on the work of Romaniello \textit{et al.} \cite{Romaniello_2012} (see also Ref.~\onlinecite{MartinBook}). Our goal is to approximate the expression of $\Sigma_\text{c}$ given in Eq.~\eqref{eq:Sig}. There are basically two ways of doing this: approximating the kernel $\Xi$ and/or the two-body correlation function $L$.

First, let us show how to recover the $GW$ form that we have derived in Sec.~\ref{sec:hedin}. In $GW$, one assumes a simple local form for the kernel by setting $\Xi = \Xi_\text{H} = - \ii V$ in Eq.~\eqref{eq:Sig}. Hence, one gets
\begin{equation}
	\Sigma_\text{c}(12) = \ii G(12) v(13) \chi(34) v(42) = \ii G(12) W_c(12)
\end{equation}
where $W_\text{c} = v \chi v$ is the correlation part of $W$.

The $GW$ approximation is obtained by setting $\Xi = \Xi_\text{H} = - \ii V$ in the expression of the propagator, which yields, thanks to Eq.~\eqref{eq:W_BSE}, the following Dyson equation for the dynamically screened Coulomb interaction 
\begin{equation}
\label{eq:Dyson_W}
\begin{split}
	W(12;1'2') 
	& = V(12;2'1') 
	\\
	& + V(13;2'3') P^\eh(34';3'4) W(42;1'4')   
\end{split}
\end{equation}
with 
\begin{equation}
\label{eq:Peh}
	P^\eh(12;1'2') = - \ii L_0(12;1'2') = - \ii G(12')G(21')
\end{equation}
the four-point version of the eh-RPA polarizability. Equations \eqref{eq:Dyson_W} and \eqref{eq:Peh} are the key equations of the $GW$ formalism and we will further discuss how to calculate these quantities in Sec.~\ref{sec:chi}.

One can also include internal and/or external vertex corrections by improving the approximation of $\Xi$ in Eqs.~\eqref{eq:BSE} and \eqref{eq:Sig}, leading to various more involved and expensive approximations.\cite{DelSol_1994,Shirley_1996,Schindlmayr_1998,Morris_2007,Shishkin_2007b,Romaniello_2009a,Romaniello_2012,Gruneis_2014,Chen_2015,Ren_2015,Caruso_2016,Hung_2017,Maggio_2017b,Cunningham_2018,Vlcek_2019,Lewis_2019a,Pavlyukh_2020,Wang_2021,Bruneval_2021,Mejuto-Zaera_2022,Wang_2022,Forster_2022b,Dadkhah_2023,Grzeszczyk_2023} 

Let us explore the derivation of the $T$-matrix self-energy from the alternative form of Hedin's equations. The main idea is to rely on a rough approximation, $L = L_0$ in Eq.~\eqref{eq:BSE}, for the response of the system but concentrate on a clever approximation of $\Xi$. In other words, one neglects the screening effects rather than the (external) vertex corrections.  To this end, we introduce an effective four-point interaction $T$, such that
\begin{equation}
	\label{eq:SigT}
	\Sigma(12) = \ii G(43) T(13;24)
\end{equation}
where, at this stage, $T$ is an unknown four-point generalized effective interaction that is linked to the kernel through the functional derivative of $\Sigma$ [see Eq.~\eqref{eq:kernel}] as follows
\begin{equation}
	\Xi(12;1'2') = \ii T(12;1'2') + \ii G(34) \fdv{T(14;23)}{G(2'2)}
\end{equation}
Additionally, we neglect the variation of $T$ with respect to $G$, i.e., $\fdv*{T}{G} = 0$, as it is usually done in the Bethe-Salpeter equation formalism, \cite{Strinati_1988} which yields $\Xi = \ii T$. Using Eqs.~\eqref{eq:SigH}, \eqref{eq:SigX}, and \eqref{eq:Sig}, the self-energy then becomes an integral equation for $T$:
\begin{equation}
\begin{split}
    \Sigma(12)
    & = \ii G(43)T(13;24)
    \\
    & = \Sigma_\text{Hx}(12) -  v(16) G(13) G(46) G(65) T(35;24)
\label{eq:S_O}
\end{split}
\end{equation}
Since $\ii G(43)T(13;24)$ cannot be directly inverted to find $T$, several choices for $T$ yield a suitable form for $\Sigma$. More explicitly, by factorizing one of the Green's functions stemming from $L_0$ or the other, i.e., $G(46)$ or $G(65)$ in Eq.~\eqref{eq:S_O}, one generates the two channels of the $T$-matrix: the particle-particle $T$-matrix, $T^\pp$, or the electron-hole $T$-matrix, $T^\eh$. (By setting $T(35;24)=-v(35)\delta(32)\delta(54)$ in the right-hand side of Eq.~\eqref{eq:S_O} and by factorizing $G(12)$, one would recover the $GW$ form of the self-energy.) More explicitly, they are defined via two distinct Dyson equations that read 
\begin{subequations}
\begin{align}
\label{eq:Dyson_Tpp}
	\begin{split}
		T^\pp(12;1'2')
		= & - \oV(12;1'2') 
		\\ 
		& + \frac{1}{4} \oV(12;34) P^\pp(34;56) T^\pp(65;1'2')
	\end{split}
	\\
\label{eq:Dyson_Teh}
	\begin{split}
		T^\eh(12;1' 2') 
		= & - \oV(12;1'2') 
		\\ 
		& - V(12';34) \overline{P}^\eh(36;45) T^\eh(52; 1'6)
	\end{split}
\end{align}
\end{subequations} 
where 
\begin{equation}
	\oV(12;1'2') = V(12;1'2') - V(12;2'1')
\end{equation}
is the four-point antisymmetrized Coulomb operator, and
\begin{subequations}
\begin{align}
\label{eq:Ppp}
	P^\pp(12;1'2') & = + \ii \qty[G(11') G(22') -  G(12') G(21')]
	\\
\label{eq:oPeh}
	\overline{P}^\eh(12;1'2') & = - \ii G(12')G(21')
\end{align} 
\end{subequations} 
are the pp-RPA \cite{Schuck_Book} and an eh-RPA-like polarizabilities, respectively. 
\titou{At this stage, the two eh-RPA polarizabilities reported in Eqs.~\eqref{eq:Peh} and \eqref{eq:oPeh} have the same expression. However, as we shall see later on, they will enter $W$ and $T^\eh$ with different spin structures.}
Note that Eq.~\eqref{eq:Dyson_Tpp} is a symmetrized version of the standard Dyson equation for $T^\pp$ given, for example, in Ref.~\onlinecite{Romaniello_2012}. It is obtained by exploiting the symmetry of the Bethe-Goldstone equation for $G_2$ with respect to the exchange of the two particles. In other words, the four terms that arise from $\oV P^\pp$ on the right-hand side of Eq.~\eqref{eq:Dyson_Tpp} are topologically equivalent, which justifies the prefactor $1/4$.

\begin{figure*}
	\includegraphics[width=0.8\linewidth]{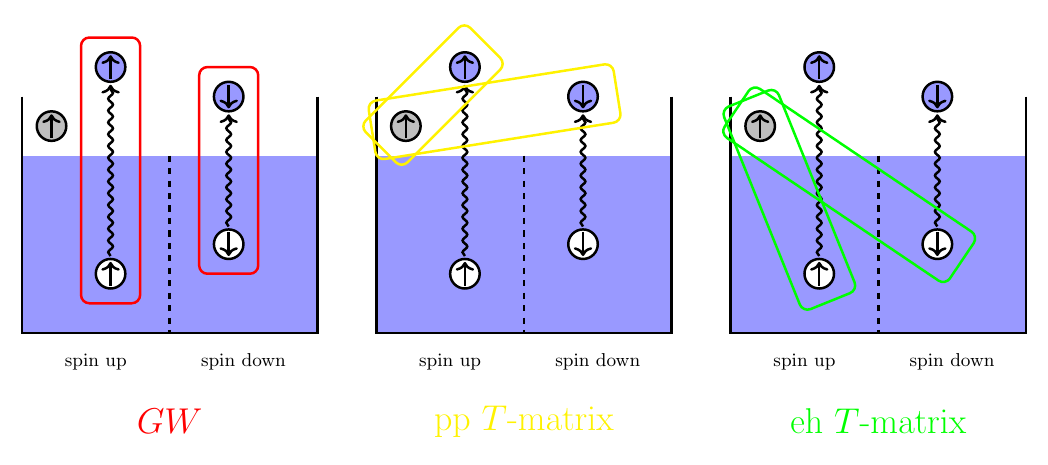}
	\caption{Schematic view of an electron attachment in the case of a closed-shell many-body system.
	The spin-up electron added to the system (gray) creates electron-hole pairs (wavy lines) in the spin-up and spin-down channels.
	The three different correlation channels that correspond to three-particle propagations are represented.
	Left: At the $GW$ level (red), the effective interaction is created by the propagation of the electron-hole pairs.
	Center: At the pp $T$-matrix level (yellow), the effective interaction is created by the propagation of the added electron and the spin-up or spin-down excited electron. 
	Right: At the eh $T$-matrix level (green), the effective interaction is created by the propagation of the added electron and the spin-up or spin-down hole.
	}
	\label{fig:dressing}
\end{figure*}

One can show that $P^\pp$ and $P^\eh$ have the same spin structure, but different time structures, while $P^\eh$ and $\overline{P}^\eh$ have the same time structure, but different spin structures (see Fig.~\ref{fig:dressing}). \cite{Romaniello_2012,MartinBook}
Equations \eqref{eq:Dyson_Tpp}, \eqref{eq:Dyson_Teh}, \eqref{eq:Ppp}, and \eqref{eq:oPeh} are the key equations of the $T$-matrix formalism, and we shall discuss in Sec.~\ref{sec:chi} how to explicitly compute their respective response functions and self-energies.

\section{Response functions}
\label{sec:chi}

As derived in Sec.~\ref{sec:dyson}, the dynamically screened Coulomb interaction $W$, Eq.~\eqref{eq:Dyson_W}, and the pp and eh $T$-matrices, Eqs.~\eqref{eq:Dyson_Tpp} and \eqref{eq:Dyson_Teh}, are given in terms of Dyson equations, with a RPA polarizability $P$ as kernel [see Eqs.~\eqref{eq:Peh}, \eqref{eq:Ppp}, and \eqref{eq:oPeh}]. However, they can be alternatively expressed in terms of a corresponding RPA response function $\chi$. This provides a formulation for the self-energy in terms of the eigenvalues and eigenvectors of the RPA matrix. This expression is textbook knowledge for $W$ and it has been recently derived for $T^\pp$. \cite{Barbieri_2007,Dickhoff_2008,Zhang_2017,Li_2021b,Li_2023} However, to the best of our knowledge, it is unknown for $T^\eh$. In the following, we provide such an expression through a derivation that puts the three approximations on an equal footing. Details regarding the derivation of $T^\eh$ are provided in the \SupMat. 

Let us explain the procedure symbolically by considering generic quantities. We start by writing a general effective interaction $\Theta$ as a function of the irreducible polarizability via a Dyson equation, i.e.,
\begin{equation}
\label{eq:Theta_1}
	\Theta = \widetilde{V} + \widetilde{V}' P \Theta
\end{equation}
where $\widetilde{V}$ and $\widetilde{V}'$ can be equal to $\pm V$ or  $\pm\oV$, and $P = P^\eh$, $P^\pp$, or $\overline{P}^\eh$. From Eq.~\eqref{eq:Theta_1}, one easily gets
\begin{equation}
\label{eq:Theta_2}
	\Theta = \epsilon^{-1} \widetilde{V}
\end{equation}
where $\epsilon = 1 - \widetilde{V}' P$ is a generalized dielectric function. Substituting the expression of $\Theta$ in the right-hand side of Eq.~\eqref{eq:Theta_1} by its expression given in Eq.~\eqref{eq:Theta_2}, we obtain the expression of the effective interaction as a function of the response function, that is,
\begin{equation}
\label{eq:Theta_3}
	\Theta = \widetilde{V} + \widetilde{V}' \chi \widetilde{V}
\end{equation}
where $\chi = P \epsilon^{-1}$, from which
\begin{equation}
\label{eq:chi}
	\chi^{-1} = P^{-1} - \widetilde{V}'
\end{equation}
This is the key equation to compute the RPA response function $\chi$ for the three channels.  
In practice, the inversion of $\chi^{-1}$ is performed by investigating the eigensystem of its matrix representation. This is the aim of Sec.~\ref{sec:Sigma}.

\section{Self-energies}
\label{sec:Sigma}
Throughout this paper, we assume real spinorbitals $\{\MO_{p}(\bx)\}$, where the composite variable $\bx = (\br,\sigma)$ gathers space and spin variables. The indices $i$, $j$, $k$, and $l$ are occupied (hole) orbitals; $a$, $b$, $c$, and $d$ are unoccupied (particle) orbitals; $p$, $q$, $r$, and $s$ indicate arbitrary orbitals; and $m$ and $n$ label single excitations/deexcitations and double electron attachments/detachments, respectively. The one-electron energies, $\{\e_{p}\}$, are quasiparticle energies and
\begin{equation}
\label{eq:ERI}
    v_{pqrs} = \iint \frac{\MO_{p}(\bx_{1})\MO_{q}(\bx_{2})\MO_{r}(\bx_{1})\MO_{s}(\bx_{2})}{\abs{\br_{1}-\br_{2}}} d\bx_{1}d\bx_{2} 
\end{equation}
are the usual bare two-electron integrals in the spinorbital basis.
For any two-electron operator $\mathcal{O}$, we follow the same convention for its projection in the spinorbital basis, i.e.,
\begin{equation}
    \mathcal{O}_{pqrs} = \iint \MO_{p}(\bx_{1})\MO_{q}(\bx_{2})\mathcal{O}(\bx_{1},\bx_{2})\MO_{r}(\bx_{1})\MO_{s}(\bx_{2}) d\bx_{1}d\bx_{2}
\end{equation}

\subsection{$GW$ self-energy}
\label{sec:GW}

As stated previously, the eh polarizability $P^\eh$ defined in Eq.~\eqref{eq:Peh} used to compute $W$ within the $GW$ approximation [see Eq.~\eqref{eq:Dyson_W}] is the usual eh-RPA polarizability where one performs a resummation of all direct ring diagrams.
The corresponding response function $\chi^\eh$ is constructed via the eigenvalues and eigenvectors of the eh-RPA linear system defined in the basis of excitations ($i \to a$) and deexcitations ($a \to i$) as follows:
\begin{multline}
\label{eq:ehRPA}
	\begin{pmatrix} 
    	\bA^\eh & \bB^\eh
		\\
    	-\bB^\eh & -\bA^\eh
	\end{pmatrix}
	\begin{pmatrix} 
    	\bX^\eh & \bY^\eh 
		\\
    	\bY^\eh & \bX^\eh 
    \end{pmatrix}
  	\\
	=
	\begin{pmatrix} 
    	\bX^\eh & \bY^\eh
		\\
    	\bY^\eh & \bX^\eh
    \end{pmatrix}
	\begin{pmatrix} 
    	\bOm^\eh & \bO
		\\
    	\bO & -\bOm^\eh
    \end{pmatrix}  
\end{multline}
where the diagonal matrix $\bOm^\eh$ gathers the positive eigenvalues and the normalization condition is
\begin{equation}
	\label{eq:eh_norm}
	\T{\begin{pmatrix} 
    	\bX^\eh & \bY^\eh 
		\\
    	\bY^\eh & \bX^\eh 
    \end{pmatrix}}
	\begin{pmatrix} 
    	\bX^\eh & \bY^\eh 
		\\
    	-\bY^\eh & -\bX^\eh 
    \end{pmatrix}
    =
	\begin{pmatrix} 
    	\bId & \bO
		\\
    	\bO & -\bId
    \end{pmatrix}    
\end{equation}
The matrix elements of the (anti)resonant block $\bA^\eh$ and the coupling block $\bB^\eh$ read 
\begin{subequations}
\begin{align}
	\label{eq:Aeh}
    A^\eh_{ia,jb} & = (\e_{a}-\e_{i}) \delta_{ij}\delta_{ab} + v_{ibaj} 
    \\
	\label{eq:Beh}
    B^\eh_{ia,jb} & =  v_{ijab}
\end{align}
\end{subequations}
Note that, in Eqs.~\eqref{eq:Aeh} and \eqref{eq:Beh}, only the direct Coulomb terms, $v_{ibaj}$ and $v_{ijab}$, are present.
Hence, the RPA eigenvalue problem \eqref{eq:ehRPA} is often referred to as \textit{direct} RPA (dRPA) in contrast to RPA with exchange (RPAx) where the corresponding exchange terms, $-v_{ibja}$ and $-v_{ijba}$, are also included. \cite{Maggio_2018}

Using these quantities, one can compute the elements of the dynamically screened Coulomb interaction as 
\begin{equation}
	W_{pqrs}(\omega)
	= v_{pqrs} 
	+ \sum_{m} \qty[ 
	\frac{M^\eh_{pr,m} M^\eh_{qs,m}}{\omega - \Om^\eh_m + \ii\eta}
	-\frac{M^\eh_{pr,m} M^\eh_{qs,m}}{\omega + \Om^\eh_m - \ii\eta}
	]
\label{W}
\end{equation}
where the screened two-electron integrals (or transition densities) read
\begin{equation}
	M_{pq,m}^\eh=\sum_{jb} v_{pjqb} \qty(X^\eh_{jb,m} + Y^\eh_{jb,m}) 
\end{equation}
and $\eta$ is a positive infinitesimal.
Performing the final convolution of the Green's function and the dynamically screened interaction, the elements of the correlation part of the $GW$ self-energy are found to be
\begin{equation}
	\label{eq:SigGW}
	\Sigma^{\eh}_{\text{c},pq}(\omega) 
	= \sum_{im}\frac{M_{pi,m}^\eh M_{qi,m}^\eh}{\omega - \e_{i} + \Om^\eh_{m} - \ii\eta}
	+\sum_{am}\frac{M_{pa,m}^\eh M_{qa,m}^\eh}{\omega - \e_{a} - \Om^\eh_{m} + \ii\eta}
\end{equation}

In the popular one-shot scheme, known as $G_0W_0$ in the case of the $GW$ approximation, \cite{Strinati_1980,Hybertsen_1985a,Godby_1988,Linden_1988,Northrup_1991,Blase_1994,Rohlfing_1995} one often considers the diagonal part of the self-energy and performs a single iteration of Hedin's equations. 
Considering a Hartree-Fock (HF) starting point, the quasiparticle energies are thus obtained by solving the non-linear quasiparticle equation for each orbital $p$:
\begin{equation}
	\omega - \e_{p}^\HF - \Re[\Sigma^{\eh}_{\text{c},pp}(\omega)] = 0
\end{equation}
It is also practically convenient to compute the renormalization factor $Z^{\eh}_{p}$ that gives the spectral weight of the corresponding quasiparticle solution $\e_{p}$: 
\begin{equation}
	\qty( Z^{\eh}_{p} )^{-1}
	= 1 - \eval{\pdv{\Re[\Sigma^{\eh}_{\text{c},pp}(\omega)]}{\omega}}_{\omega=\e_{p}}
\end{equation}
which can easily be shown to be strictly restricted between $0$ and $1$ in the case of $GW$.
When the so-called quasiparticle approximation holds, the weight of the quasiparticle equation is close to unity, while the remaining weight is distributed among the satellite (or shake-up) transitions.
\subsection{Particle-particle $T$-matrix self-energy}
\label{sec:ppGT}

The pp response function, $\chi^\pp$, is built using the eigenvalues and eigenvectors of the pp-RPA problem, a non-Hermitian eigenvalue problem expressed in the basis of double electron attachments (ee) and double electron detachments (hh): \cite{Schuck_Book,Peng_2013,Scuseria_2013}
\begin{multline}
	\begin{pmatrix}  
		\bA^\ee & \bB^{\ee,\hh} 
		\\
		-\T{(\bB^{\ee,\hh})} & -\bC^\hh
	\end{pmatrix}
	\begin{pmatrix}  
		\bX^\ee & \bY^\hh
		\\
		\bY^\ee & \bX^\hh
	\end{pmatrix}
	\\
	=
	\begin{pmatrix}  
		\bX^\ee & \bY^\hh
		\\
		\bY^\ee & \bX^\hh
	\end{pmatrix}
	\begin{pmatrix}  
		\bOm^\ee & \bO
		\\
		\bO & \bOm^\hh
	\end{pmatrix}
\end{multline}
where the diagonal matrices $\bOm^\ee$ and $\bOm^\hh$ collect the double electron attachment and double electron removal energies, and the normalization condition is 
\begin{equation}
	\T{\begin{pmatrix} 
    	\bX^\ee & \bY^\hh 
		\\
    	\bY^\ee & \bX^\hh 
    \end{pmatrix}}
	\begin{pmatrix} 
    	\bX^\ee & \bY^\hh 
		\\
    	-\bY^\ee & -\bX^\hh 
    \end{pmatrix}
    =
	\begin{pmatrix} 
    	\bId & \bO
		\\
    	\bO & -\bId
    \end{pmatrix}    
\end{equation}
The matrix elements of the different blocks are
\begin{subequations}    
\begin{align}    
    A^\ee_{ab,cd} &= (\e_{a}+\e_{b}) \delta_{ac}\delta_{bd} + \ov_{abcd}
    \\
    B^{\ee,\hh}_{ab,ij} &= \ov_{abij}
    \\   
    C^\hh_{ij,kl} &= -(\e_{i}+\e_{j}) \delta_{ik}\delta_{jl}+\ov_{ijkl}
\end{align}
\end{subequations}
where $\ov_{pqrs} = v_{pqrs} - v_{pqsr}$ are the antisymmetrized two-electron integrals.

As first derived by Zhang \textit{et al.}, the elements of the pp $T$-matrix are \cite{Zhang_2017}
\begin{equation}
	T_{pqrs}^\pp(\omega)
	= \ov_{pqrs} 
	+ \sum_{n}\qty[ 
	\frac{M_{pq,n}^\ee M_{rs,n}^\ee}{\omega - \Om^\ee_n + \ii\eta}
	- \frac{M_{pq,n}^\hh M_{rs,n}^\hh}{\omega + \Om^\hh_n - \ii\eta}
	]
\label{Tpp}
\end{equation}
where
\begin{subequations}
\begin{align}
	M_{pq,n}^\ee 
	& = \sum_{c<d}\ov_{pqcd} X^\ee_{cd,n}
	+ \sum_{k<l}\ov_{pqkl}Y^\ee_{kl,n}
	\\
	M_{pq,n}^\hh 
	& = \sum_{c<d} \ov_{pqcd} X^\hh_{cd,n} 
	+ \sum_{k<l} \ov_{pqkl} Y^\hh_{kl,n}
\end{align}
\end{subequations}
while the corresponding self-energy elements read
\begin{equation}
	\Sigma^{\pp}_{\text{c},pq}(\omega) 
	= \sum_{in}\frac{M_{pi,n}^\ee M_{qi,n}^\ee}{\omega + \e_{i} - \Om^\ee_{n} + \ii\eta}
	+ \sum_{an}\frac{M_{pa,n}^\hh M_{qa,n}^\hh}{\omega + \e_{a} - \Om^\hh_{n} - \ii\eta}
\end{equation}
with the renormalization factor fulfilling $0 \le Z^{\pp}_{p} \le 1$.
As in Sec.~\ref{sec:GW}, one denotes the one-shot scheme as $G_0T_0^\pp$.

\subsection{Electron-hole $T$-matrix self-energy}
\label{sec:ehGT}

The eh response function, $\overline{\chi}^\eh$, is obtained from a distinct RPA problem that is very similar to the usual eh-RPA problem discussed above [see Eq.~\eqref{eq:ehRPA}]. However, one has to consider index exchanges between the two coupled single (de)excitations (see Fig.~\ref{fig:dressing}). More explicitly, it reads 
\begin{multline}
	\label{eq:Casida_oTeh}
	\begin{pmatrix} 
    	\obA^\eh & \obB^\eh
		\\
    	-\obB^\eh & -\obA^\eh
	\end{pmatrix}
	\begin{pmatrix} 
    	\obX^\eh  & \obY^\eh 
		\\
    	\obY^\eh  & \obX^\eh
    \end{pmatrix}
  	\\
	=
	\begin{pmatrix} 
    	\obX^\eh  & \obY^\eh  
		\\
    	\obY^\eh  & \obX^\eh  
    \end{pmatrix}
	\begin{pmatrix} 
    	\obOm^\eh  & \bO
		\\
    	\bO & -\obOm^\eh 
    \end{pmatrix}  
\end{multline}
with a similar normalization condition as in Eq.~\eqref{eq:eh_norm}, and where
\begin{subequations}
\begin{align}
	\label{eq:oAeh}
    \oA^\eh_{ia,jb} & = (\e_{a}-\e_{i}) \delta_{ij}\delta_{ab} - v_{ibja} 
    \\
	\label{eq:oBeh}
    \oB^\eh_{ia,jb} & = - v_{ijba}
\end{align}
\end{subequations}
One would notice that it is exactly the ``exchange'' version of the usual eh-RPA problem defined in Eq.~\eqref{eq:ehRPA}. 
After a careful derivation (see the \SupMat), one eventually ends up with the following expression for the elements of the eh $T$-matrix
\begin{equation}
	\label{eq:Teh_pqrs}
	T_{pqrs}^\eh(\omega)
	= \ov_{pqrs} 
	- \sum_{m} \qty[
	\frac{L_{ps,m}^\eh R_{rq,m}^\eh}{\omega - \oOm^\eh_{m} + \ii\eta}
	- \frac{L_{sp,m}^\eh R_{qr,m}^\eh}{\omega + \oOm^\eh_{m} - \ii\eta}
	]
\end{equation}
which has the peculiarity of having numerators composed of two different sets of transition densities:
\begin{subequations}
\begin{align}
	L_{pq,m}^\eh & = \sum_{jb} \qty( v_{pjbq} \oX^\eh_{jb,m} + v_{pbjq} \oY^\eh_{jb,m} )
	\\
	R_{pq,m}^\eh & = \sum_{jb} \qty(  \ov_{pjbq} \oX^\eh_{jb,m} +  \ov_{pbjq} \oY^\eh_{jb,m} )
\end{align}
\end{subequations}
which are not symmetric with the exchange of the indices $p$ and $q$.
The resulting elements of the correlation part of the eh $T$-matrix self-energy are
\begin{equation}
\label{eq:oSigeh}
	\overline{\Sigma}^{\eh}_{\text{c},pq}(\omega) 
	= \sum_{im} \frac{L_{ip,m}^{\eh} R_{iq,m}^\eh}{\omega - \e_{i} + \oOm^\eh_m - \ii\eta}
	+ \sum_{am} \frac{L_{pa,m}^\eh R_{qa,m}^{\eh}}{\omega - \e_{a} - \oOm^\eh_m + \ii\eta}
\end{equation}
Equation \eqref{eq:oSigeh} is the central result of the present manuscript. The spin-adapted expression of the eh $T$-matrix self-energy is given in Appendix \ref{app:spin}. 
We denote the corresponding one-shot scheme as $G_0T_0^\eh$.

The renormalization factor associated with the eh $T$-matrix self-energy is expressed as follows:
\begin{equation}
	\overline{Z}^{\eh}_{p}
	= \frac{1}{1 - \eval{\pdv{\Re[\overline{\Sigma}^{\eh}_{\text{c},pp}(\omega)]}{\omega}}_{\omega=\e_{p}}}
\end{equation}
It is important to note that, unlike in $GW$ and pp $T$-matrix, $\overline{Z}^{\eh}_{p}$ is not confined within the interval of 0 to 1, and its values can extend beyond this range, including values below 0 and above 1. 
Indeed, while the values of the self-energy derivate are always positive for $GW$ and $GT^\pp$, negative values can be reached in the $GT^\eh$ formalism. This can be traced back to the eigenvectors of the eh-RPA-like matrix.
Notably, in the work by Muller \textit{et al.}, \cite{Muller_2019} it is mentioned that the spectral function of the eh $T$-matrix can assume negative values as observed in cases like iron. This phenomenon, linked to the violation of causality, directly arises due to the absence of certain self-energy diagrams. It is acknowledged that these extreme renormalization effects should be regarded as unphysical. 

In the context of solids, the eh $T$-matrix approximation is often used to study electron-magnon scattering processes in ferromagnetic systems.  \cite{Muller_2019,Myczak_2019,Friedrich_2019,Nabok_2021}
However, to the best of our knowledge, calculations of quasiparticle energies in realistic molecular systems within the eh $T$-matrix approximation have never been reported in the literature

\section{Results and discussion}
\label{sec:res}

In this study, we exclusively employ the restricted formalism due to all investigated systems possessing a closed-shell singlet ground state. Our calculations are initiated from Hartree-Fock (HF) orbitals and energies. We focus on a set composed by charged excitations where we specifically consider principal ionization potentials (IPs). This set consists of 20 atoms and molecules, known as the $GW20$ set, which is part of the $GW100$ test set \cite{vanSetten_2015} and has been previously explored in Refs.~\onlinecite{Lewis_2019a,Loos_2020,Monino_2023}. The geometries for the $GW20$ set are extracted from Ref.~\onlinecite{vanSetten_2015}. 

Using the def2-TZVPP basis, we employ the three one-shot schemes discussed in the present paper to compute the IPs: $G_0W_0$, $G_0T_0^\pp$, and $G_0T_0^\eh$. These three many-body formalisms have been implemented in \textsc{quack}, an open-source software for emerging quantum electronic structure methods, which source code is available at \url{https://github.com/pfloos/QuAcK}. For each scheme, we compute the quasiparticle energies as explained in Sec.~\ref{sec:GW} using Newton's method. As reference data, we rely on the IPs computed (in the same basis) via energy difference between the cation and the neutral species using coupled cluster singles and doubles with perturbative triples [$\Delta$CCSD(T)]. \cite{Krause_2015} Throughout our calculations, we set the positive infinitesimal $\eta$ to zero.

\begin{table*}
\caption{Principal IPs (in eV) of the $GW20$ set computed at various levels of theory using the def2-TZVPP basis.
The corresponding renormalization factor is reported in parenthesis.
The mean absolute error (MAE), mean signed error (MSE), root-mean-square error (RMSE), and maximum error (Max) with respect to the reference $\Delta$CCSD(T) values are reported.}
\label{tab:IPs_def2-tzvpp}
	\begin{ruledtabular}
		\begin{tabular}{ldddd}
Mol. 		 & \tabc{$G_0W_0$}  & \tabc{$G_0T_0^\pp$} & \tabc{$G_0T_0^\eh$} & \tabc{$\Delta$CCSD(T)} \\					
		\hline
\ce{He} 	&	24.60(0.96)	&	24.75(0.99)	&	24.26(0.91)			&	24.51	\\
\ce{Ne} 	&	21.35(0.95)	&	21.02(0.96)	&	18.69(0.83)			&	21.32	\\
\ce{H2} 	&	16.48(0.95)	&	16.26(0.99)	&	17.26(0.86)			&	16.40	\\
\ce{Li2} 	&	5.29(0.92)	&	5.04(0.98)	&	4.76(0.61)\fnm[1]	&	5.27 	\\
\ce{LiH} 	&	8.15(0.92)	&	8.14(0.98)	&	7.35(0.46)			&	7.96 	\\
\ce{HF} 	&	16.17(0.94)	&	15.65(0.95)	&	13.23(0.76)			&	16.03	\\
\ce{Ar} 	&	15.73(0.95)	&	15.52(0.97)	&	16.03(0.83)			&	15.54	\\
\ce{H2O} 	&	12.82(0.94)	&	12.28(0.95)	&	10.48(0.73)			&	12.56	\\
\ce{LiF} 	&	11.31(0.92)	&	10.88(0.94)	&	7.98(0.69)			&	11.32	\\
\ce{HCl} 	&	12.77(0.95)	&	12.50(0.96)	&	13.21(0.79)			&	12.59	\\
\ce{BeO} 	&	9.76(0.91)	&	9.20(0.93)	&	7.94(0.33)			&	9.94 	\\
\ce{CO} 	&	15.00(0.93)	&	14.44(0.95)	&	15.42(0.24)			&	14.21	\\
\ce{N2} 	&	16.30(0.93)	&	15.69(0.94)	&	14.72(0.69)			&	15.57	\\
\ce{CH4} 	&	14.74(0.94)	&	14.27(0.96)	&	14.46(0.79)			&	14.37	\\
\ce{BH3} 	&	13.64(0.94)	&	13.30(0.97)	&	13.87(0.81)			&	13.28	\\
\ce{NH3} 	&	11.14(0.94)	&	10.64(0.95)	&	9.87(0.73)			&	10.68	\\
\ce{BF} 	&	11.26(0.94)	&	10.91(0.98)	&	16.18(0.65)			&	11.09	\\
\ce{BN} 	&	11.69(0.92)	&	11.11(0.94)	&	13.29(0.14)\fnm[1]	&	11.89	\\
\ce{SH2} 	&	10.48(0.94)	&	10.17(0.96)	&	11.28(0.76)			&	10.31	\\
\ce{F2} 	&	16.27(0.93)	&	15.36(0.93)	&	11.19(0.71)\fnm[1]	&	15.71	\\
\hline
MAE 		&	0.26		&	0.25	&	1.59	\\
MSE 		&	0.22		&	-0.17	&	-0.45	\\
RMSE 		&	0.34		&	0.32	&	2.11	\\
Max 		&	0.79		&	0.78	&	5.09	\\
		\end{tabular}
	\end{ruledtabular}
	\fnt[1]{ Calculation of $T^\eh$ performed in the Tamm-Dancoff approximation due to triplet instabilities.}
\end{table*}

\begin{figure*}
	\includegraphics[width=0.8\linewidth]{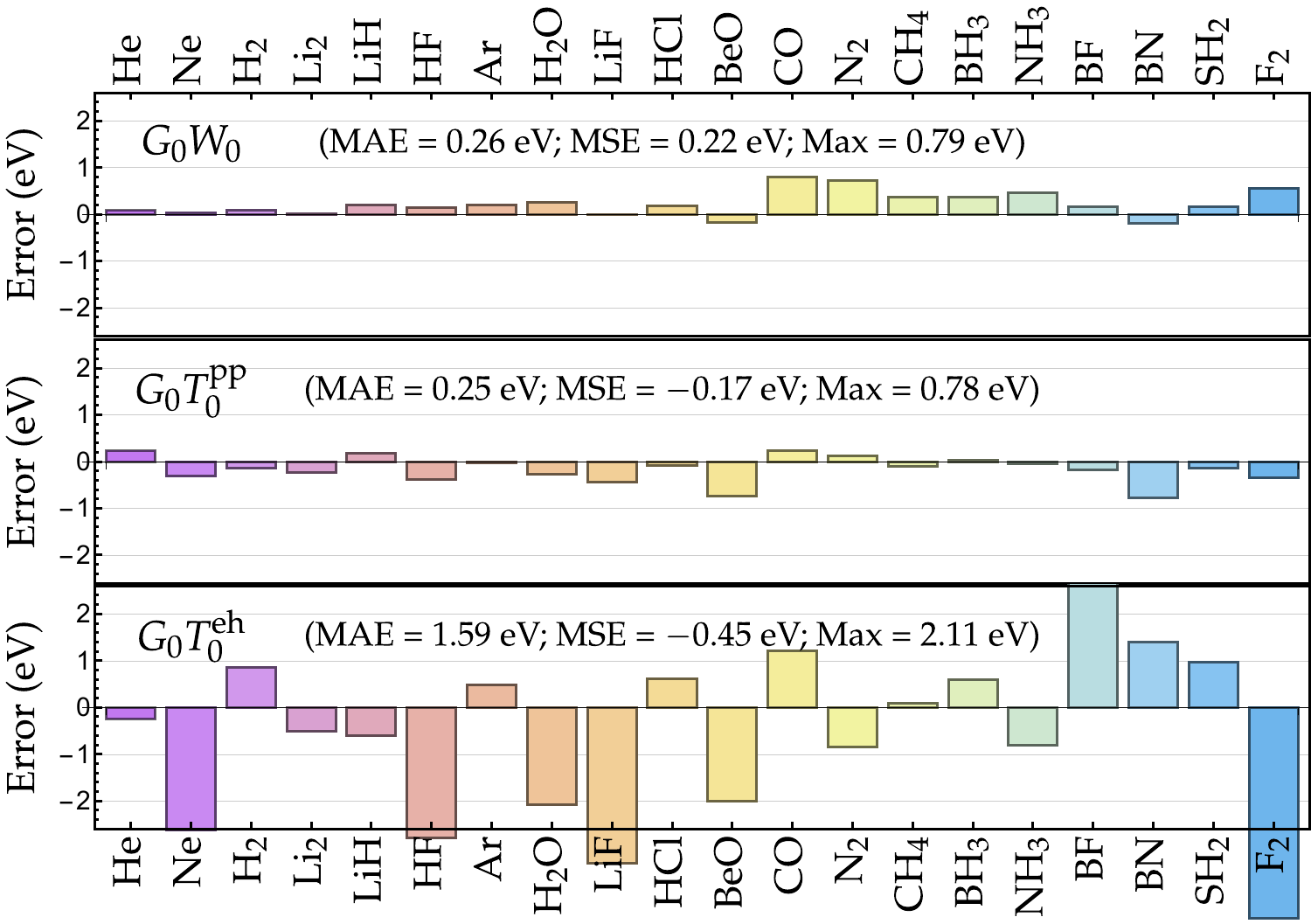}
	\caption{Error (in eV) with respect to $\Delta$CCSD(T) in the principal IPs of the $GW20$ set computed at the $G_0W_0$, $G_0T_0^\pp$, and $G_0T_0^\eh$ levels with the def2-TVZPP basis. Raw data can be found in Table \ref{tab:IPs_def2-tzvpp}.
	}
	\label{fig:IPs_def2-tzvpp}
\end{figure*}

The principal IPs of the $GW20$ test are reported in Table \ref{tab:IPs_def2-tzvpp} and the error with respect to the $\Delta$CCSD(T) reference values are represented in Fig.~\ref{fig:IPs_def2-tzvpp}. As previously reported in the literature, \cite{Monino_2023} $G_0W_0$ and $G_0T_0^\pp$ have very similar mean absolute errors (MAEs) for this set of small systems (\SI{0.26}{\eV} \textit{vs} \SI{0.26}{\eV}), while their respective mean signed errors (MSEs) are almost exactly opposite (\SI{0.22}{\eV} \textit{vs} \SI{-0.17}{\eV}). The $G_0T_0^\eh$ scheme has much larger MAE (\SI{1.59}{\eV}) and MSE (\SI{-0.45}{\eV}). Figure \ref{fig:IPs_def2-tzvpp} clearly shows that large errors are observed for some systems, like \ce{Ne}, \ce{HF}, \ce{LiF}, \ce{BeO}, \ce{BF}, and \ce{F2}. Moreover, even at equilibrium geometry, triplet instabilities are encountered for several systems (\ce{Li2}, \ce{BN}, and \ce{F2}). This forced us to compute $T^\eh$ within the Tamm-Dancoff approximation which consists in setting $\obB = \bO$ in Eq.~\eqref{eq:Casida_oTeh}. From these results, it is clear that the performances of $G_0T_0^\eh$ are clearly inferior to those of $G_0W_0$ and $G_0T_0^\pp$. This explains the development of the screened version of the eh $T$-matrix in solid-state calculations. \cite{Springer_1998,Romaniello_2012,Zhukov_2004,Zhukov_2005,Zhukov_2006,Nechaev_2006,Nechaev_2008,Monnich_2006} Qualitatively at least, the poor performance of $G_0T_0^\eh$ can be explained by the fact that $T^\eh$ is constructed with the eigenvectors and eigenvalues associated with the triplet states of the system computed at the RPAx level [see Eq.~\eqref{eq:Casida_oTeh} and the discussion below it] starting from a singlet HF ground-state reference. It is well known that this usually leads to poorly described triplet states and, often, triplet instabilities \cite{Dreuw_2005} (see Appendix \ref{app:spin}).

\begin{figure*}
	\includegraphics[width=0.45\linewidth]{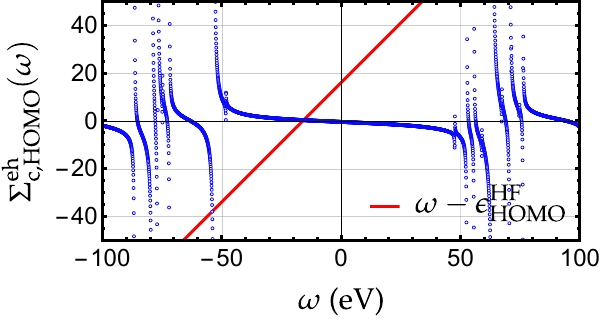}
	\hspace{0.08\linewidth}
	\includegraphics[width=0.45\linewidth]{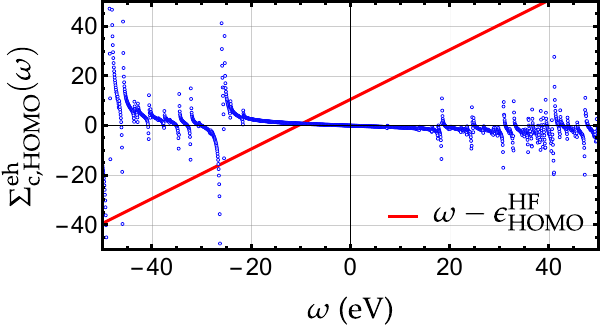}
	\\
	\includegraphics[width=0.45\linewidth]{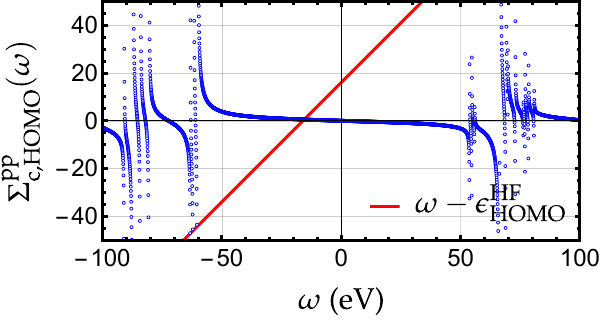}
	\hspace{0.08\linewidth}
	\includegraphics[width=0.45\linewidth]{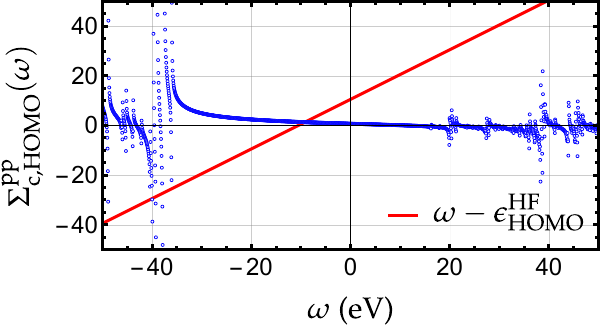}
	\\
	\includegraphics[width=0.45\linewidth]{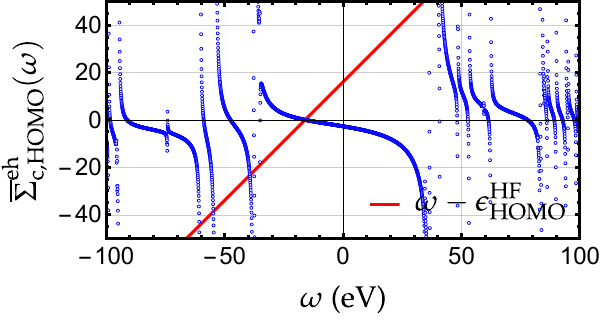}
	\hspace{0.08\linewidth}
	\includegraphics[width=0.45\linewidth]{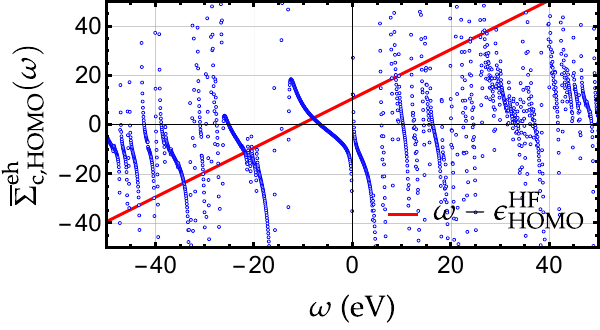}
	\caption{Self-energy (blue curves) associated with the HOMO of \ce{Ar} (left) and \ce{BeO} (right) computed at the $G_0W_0$ (top), $G_0T_0^\pp$ (middle), and $G_0T_0^\eh$ (bottom) levels with the def2-TVZPP basis. The solutions of the quasiparticle equation are given by the intersection of the blue and red curves. Raw data can be found in {\SupMat}.}
	\label{fig:SigC}
\end{figure*}

In Fig.~\ref{fig:SigC}, we plot the variation of the $G_0W_0$ (top), $G_0T_0^\pp$ (middle), and $G_0T_0^\eh$ (bottom) self-energies associated with the highest-occupied molecular orbital (HOMO) as functions of $\omega$ \titou{for two systems with weakly correlated $N$-electron ground state, namely, \ce{Ar} and \ce{BeO}}. The solutions of the quasiparticle equation are given by the intersection of the blue and red curves. At the $G_0W_0$ and $G_0T_0^\pp$ levels, the two systems exhibit similar behavior with well-defined quasiparticle solutions with respective weights of $0.95$ ($0.98$) and $0.97$ ($0.93$) for \ce{Ar} (\ce{BeO}), as reported in Table \ref{tab:IPs_def2-tzvpp}. This is graphically evidenced by the small values of the self-energy derivative in the central region of the graphs.
At the $G_0T_0^\eh$ level, it is clear that the variations of the self-energy are more pronounced.
Contrary to $G_0W_0$ and $G_0T_0^\pp$, the $G_0T_0^\eh$ self-energy derivative can take positive values, as mentioned in Sec.~\ref{sec:ehGT}. For \ce{Ar}, the quasiparticle solution has a weight of $0.83$ and the behavior of the $G_0T_0^\eh$ self-energy is quite standard. The case of \ce{BeO} is more interesting though as the solution around \SI{-8}{\eV} reached from the HF starting value using Newton's method has a small weight ($0.33$) and cannot really be classified as a quasiparticle solution. Another solution with a similar weight can be located around \SI{-22}{\eV}. This example represents a clear breakdown of the quasiparticle approximation.

\begin{figure*}
	\includegraphics[width=\linewidth]{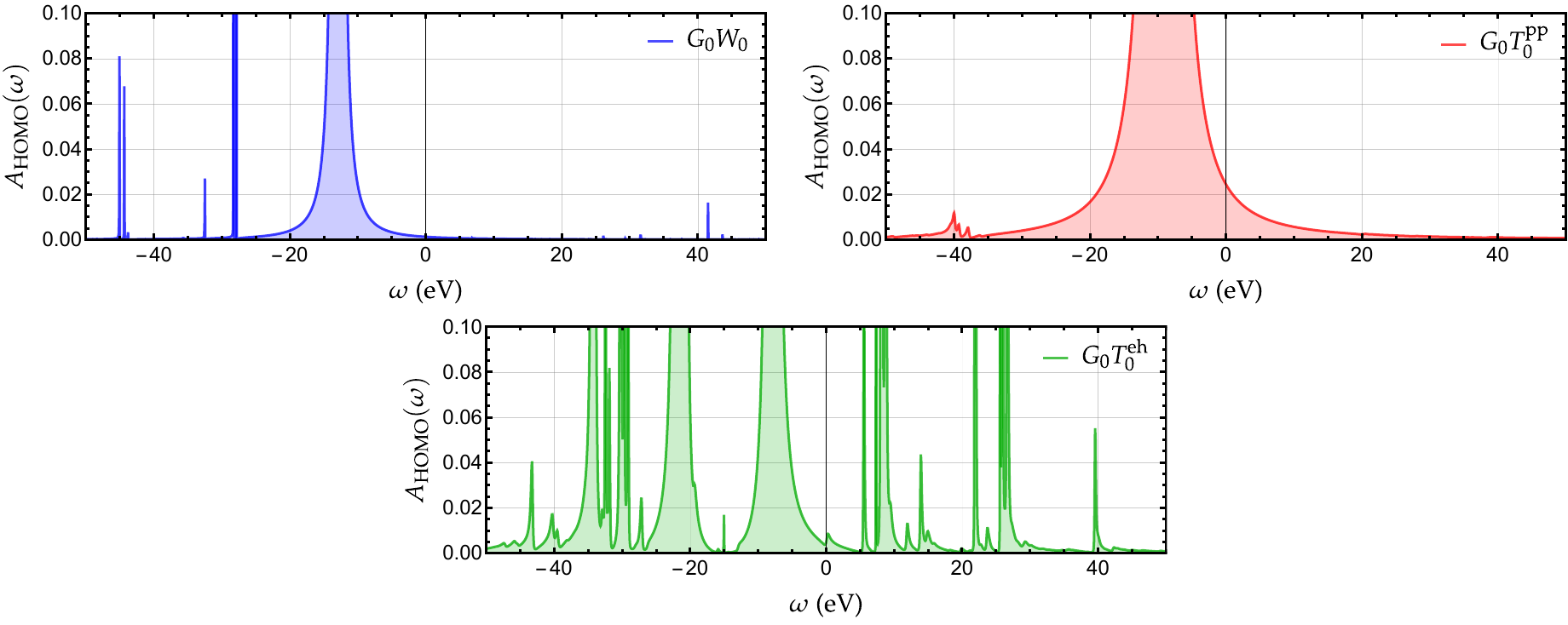}
	\caption{\titou{Spectral function associated with the HOMO of \ce{BeO} computed at the $G_0W_0$ (top left), $G_0T_0^\pp$ (top right), and $G_0T_0^\eh$ (bottom) levels with the def2-TVZPP basis with $\eta = \SI{0.01}{\hartree}$. Raw data can be found in {\SupMat}.}}
	\label{fig:A}
\end{figure*}

\titou{This distribution of spectral weight amongst several solutions can be analyzed in more detail by looking at the spectral function associated with each orbital $p$, which, as readily seen from the following equation, is related to the imaginary part of the one-body Green's function:
\begin{equation}
\label{eq:spectral_function}
\begin{split}
	A_{p}(\omega) 
	& = \frac{1}{\pi} \abs{ \Im G_{pp}(\omega)} 
	\\
	& = \frac{\eta/\pi}{\qty[\omega - \e_p^{\HF} - \Sigma_{\text{c},pp}(\omega)]^2 + \eta^2}
\end{split}
\end{equation}
The spectral function associated with the HOMO of \ce{BeO} computed at the three levels of theory is depicted in Fig.~\ref{fig:A}.
While there is a well-defined quasiparticle peak with $G_0W_0$ and $G_0T_0^\pp$, it is clear that there are two main solutions for $G_0T_0^\eh$ surrounded by a large number of satellite peaks.}

\titou{Next, we consider more strongly correlated systems, namely, \ce{B2} and \ce{C2}, \cite{Bauschlicher_1987,Bruna_1990,Abrams_2004,Sherrill_2005,Li_2006,Booth_2011} at their respective experimental equilibrium geometry: $R_{\ce{B-B}} = \SI{1.5900}{\angstrom}$ and $R_{\ce{C-C}} = \SI{1.2425}{\angstrom}$. \cite{HerzbergBook}
Based on full configuration interaction (FCI) calculations, performed with \textsc{quantum package}, \cite{Garniron_2019} both the neutral and charged species (cation and anion) have strong multireference characters with several determinants contributing to their respective ground-state wave functions. In the case of \ce{B2}, the FCI calculations show that the weight of the HF determinant is only $0.36$ with another doubly-excited determinant of weight $0.33$. Likewise, in the carbon dimer, the weight of the HF determinant is $0.69$ with another doubly-excited determinant of weight $0.1$. The cationic and anionic states of these systems are also of multireference character, although not as much as \ce{B2}.}

\titou{The values of the principal IP computed at the $G_0W_0$, $G_0T_0^\pp$, and $G_0T_0^\eh$ levels of theory with the def2-TZVPP basis are reported in Table \ref{tab:res_MR} for these two strongly correlated systems. The FCI values are also reported for comparison purposes. Note that, here again, we rely on a restricted HF starting point. It is worth noting that, because of triplet instabilities, the calculations of $T^\eh$ are performed in the Tamm-Dancoff approximation.}

\titou{For the boron dimer, the IP calculated at the $G_0T_0^\eh$ level is remarkably close at \SI{8.91}{\eV}, deviating by only \SI{0.06}{\eV} from the FCI value of \SI{8.97}{\eV.} In this case, $G_0W_0$ (at \SI{9.06}{\eV}) and $G_0T_0^\pp$ (at \SI{8.69}{\eV}) are also in proximity to the FCI result. The situation takes a somewhat different turn for \ce{C2}, where $G_0T_0^\pp$ (at \SI{12.63}{\eV}) closely approximates the FCI value of \SI{12.45}{\eV}, while $G_0W_0$ (at \SI{12.92}{\eV}) and $G_0T_0^\eh$ (at \SI{12.63}{\eV}) exhibit slightly larger deviations, on the order of half an eV. As for the systems composing the $GW20$ set (see above), the spectral weight of the quasiparticle solution computed at the $G_0T_0^\eh$ level is usually significantly lower than the values obtained from $G_0W_0$ and $G_0T_0^\pp$.}

\begin{table}
\caption{\titou{Principal IP (in \si{\eV}) for \ce{B2} and \ce{C2} computed at the $G_0W_0$, $G_0T_0^\pp$, $G_0T_0^\eh$, and FCI levels of theory with the def2-TZVPP basis. The corresponding renormalization factor is reported in parenthesis.}}
\label{tab:res_MR}
	\begin{ruledtabular}
		\begin{tabular}{lrrrr}
Mol. 	 	& \tabc{$G_0W_0$} 		& \tabc{$G_0T_0^\pp$} 	& \tabc{$G_0T_0^\eh$\fnm[1]}	&	\tabc{FCI} \\					
		\hline
\ce{B2}		&	9.06(0.93)	&	8.69(0.96)	&	8.91(0.75)	&	8.97	\\
\ce{C2}		&	12.92(0.93)	&	12.63(0.95)	&	13.01(0.76)	&	12.45	\\
		\end{tabular}
	\end{ruledtabular}
	\fnt[1]{Calculation of $T^\eh$ performed in the Tamm-Dancoff approximation due to triplet instabilities.}
\end{table}

\section{Concluding remarks}
\label{sec:ccl}
This manuscript presents a comprehensive derivation and implementation of the explicit expressions for the three self-energies intrinsic to many-body perturbation theory: the well-established $GW$ self-energy, along with the pp and eh $T$-matrix self-energies. To evaluate the efficacy of these approaches, we assess their performances on molecular systems. Specifically, we compute the principal IPs across a collection of 20 small molecules. 

The outcomes of our computations distinctly indicate that the eh $T$-matrix formalism falls short when compared to the other two approaches. The subset of diagrams composed by the eh ladder diagrams is thus less relevant than the two other subsets (direct rings and pp ladders) in the present context. This observation paves the way for an investigation into the screened version of the eh $T$-matrix, which has demonstrated successes in diverse systems, such as ferromagnetic periodic structures, as reported in prior studies. \cite{Muller_2019,Myczak_2019,Friedrich_2019,Nabok_2021} Another avenue for further exploration involves the combination of these three correlation channels, akin to ``fluctuation exchange'' (FLEX), \cite{Bickers_1989a,Bickers_1989b,Bickers_1991,Bickers_2004} the Baym-Kadanoff approximation, \cite{Baym_1961,Baym_1962} parquet theory, \cite{DeDominicis_1964a,DeDominicis_1964b} and other similar approaches. \cite{Romaniello_2012,Shepherd_2014,Nabok_2021,Hofierka_2022,Riva_2022,Riva_2023} Although challenging, this task holds significant promise and represents a potential avenue for our future investigations.

\titou{We have also investigated the performances of $G_0W_0$, $G_0T_0^\pp$, and $G_0T_0^\eh$ for more strongly correlated systems, namely, the boron and carbon dimers. In this context, the eh $T$-matrix scheme might be a more adequate approximation, yielding IPs and spectral weights in good agreement with the reference FCI values. We hope to report further on this in the near future.}

\section*{Supplementary Material}
\label{sec:supmat}
See the supplementary material for a detailed derivation of the eh $T$-matrix elements and the corresponding self-energy elements.

\acknowledgments{
This project has received financial support from the Centre National de la Recherche Scientifique (CNRS) through the 80$\vert$PRIME program and has been supported through the EUR grant NanoX under Grant No.~ANR-17-EURE-0009 in the framework of the ``Programme des Investissements d’Avenir''.
PFL also thanks the European Research Council (ERC) under the European Union's Horizon 2020 research and innovation programme (Grant agreement No.~863481) for funding.}

\section*{Data availability statement}
The data that supports the findings of this study are available within the article and its supplementary material.

\appendix

\section{Hedin's equations from the equation-of-motion formalism}
\label{app:hedin_alt}
Here, we present a different derivation of the standard (see Sec.~\ref{sec:hedin}) and alternative (see Sec.~\ref{sec:hedin_bis}) forms of Hedin's equations. We refer the interested reader to Refs.~\onlinecite{Romaniello_2012,MartinBook} for additional details.

From the equation-of-motion of the one-body Green's function \cite{MartinBook}
\begin{equation}
	G(12) = G_0(12) - \ii G_0(13) v(34) G_2(34^+;24^{++})
\end{equation} 
and the Dyson equation \eqref{eq:G} linking $G_0$ and $G$, one gets the following exact expression of the self-energy
\begin{equation}
	\Sigma(12) = - \ii v(1^+3) G_2(13;43^+) G^{-1}(42) 
\end{equation}
where $G_2$ is the two-body Green's function.
One can then employ the Martin-Schwinger relation \cite{Martin_1959} 
\begin{equation}
\begin{split}
	\fdv{G(12;[V_\text{ext}])}{V_\text{ext}(3)} 
	& = - G_2(13;23^+;[V_\text{ext}]) 
	\\
	& + G(12;[V_\text{ext}]) G(33^+;[V_\text{ext}])
\end{split}
\end{equation}
which relates the one- and two-body Green's functions and the variation of $G$ with respect to a fictitious external potential $V_\text{ext}$, to substitute $G_2$ in the expression of the self-energy. The equilibrium Green's functions are retrieved for $V_\text{ext}=0$, i.e., $G(12,[V_\text{ext}=0]) \equiv G(12)$ and $G_2(12;34;[V_\text{ext}]=0) \equiv G_2(12;34)$. We hence arrive at
\begin{equation}
	\Sigma(12) = \Sigma_\text{H}(12) + \ii v(1^+3)\left. \fdv{G(14;[V_\text{ext}])}{V_\text{ext}(3)}\right|_{V_\text{ext}=0} G^{-1}(42)
\end{equation}
For notational convenience, in the following, we drop the functional dependence on $V_\text{ext}$ and the limit $V_\text{ext}=0$.
Using the chain-rule derivative
\begin{equation}
	\fdv{G(12)}{V_\text{ext}(3)}= - G(14) \fdv{G^{-1}(45)}{V_\text{ext}(3)} G(52)
\end{equation}
we finally obtain
\begin{equation}
	\Sigma(12) = \Sigma_\text{H}(12) - \ii v(1^+3) G(14) \fdv{G^{-1}(42)}{V_\text{ext}(3)}
\end{equation}
where the second term of the right-hand side corresponds to the exchange-correlation part of the self-energy. 

One can then recover Hedin's form of $\Sigma_\text{xc}$ [see Eq.~\eqref{eq:Sigma_xc}] by introducing the total classical potential $V_\text{tot} = V_\text{H} + V_\text{ext}$ [where $V_\text{H}(1) = - \ii v(1^+2)G(22^+)$ is the local Hartree potential], as follows
\begin{equation}
\label{eq:Sigma_app}
\begin{split}
	\Sigma_\text{xc}(12) 
	& = - \ii v(1^+3) G(14) \fdv{G^{-1}(42)}{V_\text{tot}(5)} \fdv{V_\text{tot}(5)}{V_\text{ext}(3)}
	\\
	& = \ii G(14) W(13) \Gamma(423)
\end{split}
\end{equation}
where $W(12) = \epsilon^{-1}(13) v(32)$ is the dynamically screened Coulomb interaction, 
\begin{equation}
	\epsilon^{-1}(12) = \fdv{V_\text{tot}(1)}{V_\text{ext}(2)} = \delta(12) + v(13)\chi(32)
\end{equation}
is the inverse dielectric function [with $\chi(32) = - \ii \fdv*{G(33^+)}{V_\text{ext}(2)}$], and $\Gamma(423) = - \fdv*{G^{-1}(42)}{V_\text{tot}(3)}$ is the \textit{irreducible} vertex function. \cite{Romaniello_2012,MartinBook}

To recover the alternative form of the self-energy given in Eq.~\eqref{eq:Sig}, we substitute $G^{-1} = G_0^{-1} - V_\text{ext} - \Sigma$  into Eq.~\eqref{eq:Sigma_app}, and this yields
\begin{equation}
\begin{split}
	\Sigma_\text{xc}(12) 
	& = \Sigma_\text{x}(12) + \ii v(13) G(14) \fdv{\Sigma(42)}{G(65)} \fdv{G(65)}{V_\text{ext}(3)}
	\\
	& = \Sigma_\text{x}(12) + \ii v(13) G(14) \Xi(45;26) L(63;53)
\end{split}
\end{equation}
with the kernel $\Xi$ given by Eq.~\eqref{eq:kernel} and the polarization propagator $L(12;32) = \fdv*{G(13)}{V_\text{ext}(2)}$.

\section{Spin adaptation}
\label{app:spin}
Spin adaptation corresponds to converting spinorbital expressions to their spatial orbital equivalents for the various spin manifolds that one can encounter.
Practically, in the present case, it means that, instead of solving a larger eigenvalue problem in the spinorbital basis such as in Eq.~\eqref{eq:ehRPA}, for example, one can equivalently solve two distinct eigenvalue problems for the singlet ($2S+1=1$) and triplet ($2S+1=3$) manifolds separately:
\begin{equation}
	\begin{pmatrix}
		{}^{1,3}\bA^{\eh}		&	{}^{1,3}\bB^{\eh}	\\
		-{}^{1,3}\bB^{\eh}	&	-{}^{1,3}\bA^{\eh}	\\
	\end{pmatrix}
	\begin{pmatrix}
		{}^{1,3}\bX_{m}^{\eh}	\\
		{}^{1,3}\bY_{m}^{\eh}	\\
	\end{pmatrix}
	=
	{}^{1,3}\Om_{m}^{\eh}
	\begin{pmatrix}
		{}^{1,3}\bX_{m}^{\eh}	\\
		{}^{1,3}\bY_{m}^{\eh}	\\
	\end{pmatrix}
\end{equation}

Below, we report the spin adaptation of the eh and pp eigenvalue problems in the case of RPA, restricting ourselves to spin-restricted closed-shell calculations.
Note that, in the present appendix, although we retain the same notations as in the spinorbital basis, all the integrals and operators are defined in the spatial orbital basis.

\subsection{Electron-Hole Channel}
\label{app:eh}
Spin-adaptation of the eh channel is well documented (see, for example, Ref.~\onlinecite{Angyan_2011}). 
Each eh matrix has a similar spin structure that is directly inherited from the spin structure of the two-electron integrals.
For example, we have
\begin{equation}
	\bA^{\eh} = 
	\begin{pmatrix}
		\bA_{\upup,\upup}^{\eh}	&	\bA_{\upup,\dwdw}^{\eh}	&	\bO						&	\bO						\\
		\bA_{\dwdw,\upup}^{\eh}	&	\bA_{\dwdw,\dwdw}^{\eh}	&	\bO						&	\bO						\\
		\bO						&	\bO						&	\bA_{\updw,\updw}^{\eh}	&	\bA_{\updw,\dwup}^{\eh}	\\
		\bO						&	\bO						&	\bA_{\dwup,\updw}^{\eh}	&	\bA_{\dwup,\dwup}^{\eh}	\\
	\end{pmatrix}
\end{equation}

In the eh channel, singlet and triplet spin-adapted operators are defined as
\begin{subequations}
\begin{align}
	^{1}E_{i}^{a} & = \frac{\cre{a_\up}\ani{i_\dw} + \cre{a_\dw}\ani{i_\up}}{\sqrt{2}}
	\\
	^{3,-1}E_{i}^{a} & = \cre{a_\dw}\ani{i_\dw}
	\\
	^{3,0}E_{i}^{a} & = \frac{\cre{a_\up}\ani{i_\dw} - \cre{a_\dw}\ani{i_\up}}{\sqrt{2}}
	\\
	^{3,1}E_{i}^{a} & = \cre{a_\up}\ani{i_\up}
\end{align}
\end{subequations}
where $\cre{a_\sigma}$ ($\ani{i_\sigma}$) creates a spin-$\sigma$ electron (hole) in the virtual (occupied) orbital $a$ ($i$).
Hence, one can spin-adapt each eh matrix via the following orthogonal transformation
\begin{equation}
	\bU^{\eh} = 
	\frac{1}{\sqrt{2}}
	\begin{pmatrix}
		\bId	&	\bId	&	\bO		&	\bO		\\
		\bId	&	-\bId	&	\bO		&	\bO		\\
		\bO		&	\bO		&	\bId	&	\bId		\\
		\bO		&	\bO		&	\bId	&	-\bId	\\
	\end{pmatrix}
\end{equation}
which block-diagonalizes each eh matrix as follows:
\begin{equation}
	\T{(\bU^{\eh})}  \bA^{\eh}  \bU^{\eh} = 
	\begin{pmatrix}
		{}^{1}\bA^{\eh}		&	\bO					&	\bO					&	\bO					\\
		\bO					&	{}^{3,-1}\bA^{\eh}	&	\bO					&	\bO					\\
		\bO					&	\bO					&	{}^{3,0}\bA^{\eh}	&	\bO					\\
		\bO					&	\bO					&	\bO					&	{}^{3,1}\bA^{\eh}	\\
	\end{pmatrix}
\end{equation}
with
\begin{subequations}
\begin{align}
	\begin{split}
	{}^{1}A_{ia,jb}^{\eh} 
	= \frac{1}{2} \Big( A_{i_\up a_\up,j_\up b_\up}^{\eh} 
	& + A_{i_\up a_\up,j_\dw b_\dw}^{\eh} 
	\\
	& + A_{i_\dw a_\dw,j_\up b_\up}^{\eh} 
	+ A_{i_\dw a_\dw,j_\dw b_\dw}^{\eh} \Big)
	\end{split}
	\\
	\begin{split}
	{}^{3,0}A_{ia,jb}^{\eh} 
	= \frac{1}{2} \Big( A_{i_\up a_\up,j_\up b_\up}^{\eh} 
	& - A_{i_\up a_\up,j_\dw b_\dw}^{\eh} 
	\\
	& - A_{i_\dw a_\dw,j_\up b_\up}^{\eh} 
	+ A_{i_\dw a_\dw,j_\dw b_\dw}^{\eh} \Big)
	\end{split}
	\\
	\begin{split}
	{}^{3,\pm1}A_{ia,jb}^{\eh} 
	= \frac{1}{2} \Big( A_{i_\up a_\dw,j_\up b_\dw}^{\eh} 
	& \pm A_{i_\up a_\dw,j_\dw b_\up}^{\eh} 
	\\
	& \pm A_{i_\dw a_\up,j_\up b_\dw}^{\eh} 
	+ A_{i_\dw a_\up,j_\dw b_\up}^{\eh} \Big)
	\end{split}
\end{align}
\end{subequations}

In the case of the $GW$ approximation, using Eqs.~\eqref{eq:Aeh} and \eqref{eq:Beh}, one finds that the singlet eh-RPA matrix elements are then
\begin{subequations}
\begin{align}
	\label{eq:1A_ehRPA}
	{}^1A_{ia,jb}^{\eh} & = \delta_{ij} \delta_{ab} (\e_{a} - \e_{i}) + 2 v_{ibaj}
	\\ 
	\label{eq:1B_ehRPA}
	{}^1B_{ia,jb}^{\eh} & = 2 v_{ijab}
\end{align}
\end{subequations}
while, for the triplet manifold, it yields three identical sets of equations with matrix elements
\begin{subequations}
\begin{align}
	\label{eq:3A_ehRPA}
	{}^3A_{ia,jb}^{\eh} & = \delta_{ij} \delta_{ab} (\e_{a} - \e_{i}) 
	\\ 
	\label{eq:3B_ehRPA}
	{}^3B_{ia,jb}^{\eh} & = 0 
\end{align}
\end{subequations}
where one notes that the direct term has vanished.
Because of the form of $\chi^\eh$, the triplet excited states do not contribute to the dynamical screening within the $GW$ approximation. \cite{Monino_2021}
Hence, one only needs to solve the singlet eh-RPA problem whose elements are defined in Eqs.~\eqref{eq:1A_ehRPA} and \eqref{eq:1B_ehRPA}. 

For the eh $T$-matrix approximation, using Eqs.~\eqref{eq:oAeh} and \eqref{eq:oBeh}, one easily shows that the singlet and triplet eh-RPA matrix elements are identical:
\begin{subequations}
\begin{align}
	\label{eq:1oA_ehRPA}
	{}^1\oA_{ia,jb}^{\eh} & = {}^3\oA_{ia,jb}^{\eh} = \delta_{ij} \delta_{ab} (\e_{a} - \e_{i}) - v_{ibja}
	\\ 
	\label{eq:1oB_ehRPA}
	{}^1\oB_{ia,jb}^{\eh} & = {}^3\oB_{ia,jb}^{\eh} = - v_{ijba} 
\end{align}
\end{subequations}
which means that one only needs to solve a unique set of eigenvalue equations to compute $\overline{\chi}^\eh$. 
After spin integration, one can show that the elements of the eh $T$-matrix self-energy remain unchanged (except for the summation ranges) while the transition densities read
\begin{subequations}
\begin{align}
	L_{pq,m}^\eh & = \sum_{jb} \qty( v_{pjbq} \oX^\eh_{jb,m} + v_{pbjq} \oY^\eh_{jb,m} )
	\\
	R_{pq,m}^\eh & = \sum_{jb} \qty( \overline{\overline{v}}_{pjbq} \oX^\eh_{jb,m} + \overline{\overline{v}}_{pbjq} \oY^\eh_{jb,m} )
\end{align}
\end{subequations}
with $\overline{\overline{v}}_{pqrs} = 2 v_{pqrs} - v_{pqsr}$.

\subsection{Particle-Particle Channel}
\label{app:pp}
The spin structure of the pp problem is slightly simpler than its eh counterpart described in Appendix \ref{app:eh}.
For example, we have
\begin{equation}
	\label{eq:Aee}
	\bA^{\ee} = 
	\begin{pmatrix}
		\bA_{\upup,\upup}^{\ee}	&	\bO						&	\bO						&	\bO						\\
		\bO						&	\bA_{\dwdw,\dwdw}^{\ee}	&	\bO						&	\bO						\\
		\bO						&	\bO						&	\bA_{\updw,\updw}^{\ee}	&	\bA_{\updw,\dwup}^{\ee}	\\
		\bO						&	\bO						&	\bA_{\dwup,\updw}^{\ee}	&	\bA_{\dwup,\dwup}^{\ee}	\\
	\end{pmatrix}
\end{equation}

In the pp channel, singlet and triplet spin-adapted operators for the 2e configurations are constructed as
\begin{subequations}
\begin{align}
	^{1}E^{ab} & = \frac{\cre{b_\up}\cre{a_\dw} + \cre{b_\dw}\cre{a_\up}}{\sqrt{2(1+\delta_{ab})}} 
	\\
	^{3,-1}E^{ab} & = \cre{b_\dw} \cre{a_\dw}
	\\
	^{3,0}E^{ab} & = \frac{ \cre{b_\up}\cre{a_\dw} - \cre{b_\dw}\cre{a_\up}}{\sqrt{2}}
	\\
	^{3,1}E^{ab} & = \cre{b_\up}\cre{a_\up}
\end{align}
\end{subequations}
Similarly, for the 2h configurations, we have
\begin{subequations}
\begin{align}
	^{1}E_{ij} & = \frac{\ani{i_\up}\ani{j_\dw} + \ani{i_\dw}\ani{j_\up}}{\sqrt{2(1+\delta_{ij})}} 
	\\
	^{3,-1}E_{ij} & = \cre{i_\dw} \ani{j_\dw}
	\\
	^{3,0}E_{ij} & = \frac{\ani{i_\up}\ani{j_\dw} - \ani{i_\dw}\ani{j_\up}}{\sqrt{2}}
	\\
	^{3,1}E_{ij} & = \ani{i_\up}\ani{j_\up}
\end{align}
\end{subequations}

Hence, one can spin-adapt each matrix via the following orthogonal transformation
\begin{equation}
	\bU^{\pp} = 
	\frac{1}{\sqrt{2}}
	\begin{pmatrix}
		\bId	&	\bO		&	\bO	&	\bO		\\
		\bO	&	\bId		&	\bO	&	\bO		\\
		\bO	&	\bO		&	\bId	&	\bId		\\
		\bO	&	\bO		&	\bId	&	-\bId	\\
	\end{pmatrix}
\end{equation}
which block-diagonalizes each pp matrix as follows:
\begin{equation}
	\T{(\bU^{\pp})}  \bA^{\ee}  \bU^{\pp} = 
	\begin{pmatrix}
		{}^{1}\bA^{\ee}	&	\bO					&	\bO					&	\bO					\\
		\bO					&	{}^{3,0}\bA^{\ee}	&	\bO					&	\bO					\\
		\bO					&	\bO					&	{}^{3,-1}\bA^{\ee}	&	\bO					\\
		\bO					&	\bO					&	\bO					&	{}^{3,1}\bA^{\ee}	\\
	\end{pmatrix}
\end{equation}
with
\begin{subequations}
\begin{align}
	{}^{1}A_{ab,cd}^{\ee} 
	& = \frac{A_{a_\up b_\dw,c_\up d_\dw}^{\ee} + A_{a_\dw b_\up,c_\dw d_\up}^{\pp}}{\sqrt{1+\delta_{ab}}\sqrt{1+\delta_{cd}}} 
	\\
	{}^{3,-1}A_{ab,cd}^{\ee} 
	& = A_{a_\dw b_\dw,c_\dw d_\dw}^{\ee} 
	\\
	{}^{3,0}A_{ia,jb}^{\ee} 
	& = A_{a_\up b_\dw,c_\up d_\dw}^{\ee} - A_{a_\dw b_\up,c_\dw d_\up}^{\ee}
	\\
	{}^{3,1}A_{ab,cd}^{\pp} 
	& = A_{a_\up b_\up,c_\up d_\up}^{\ee} 
\end{align}
\end{subequations}
and similar expressions for the matrices $\bB^{\ee,\hh}$ and $\bC^{\hh}$.

The singlet pp-RPA matrix elements are then
\begin{subequations}
\begin{align}
	\label{eq:1A_ppRPA}
	{}^1A_{ab,cd}^{\ee} 
	& = \delta_{ac} \delta_{bd} (\e_{a} + \e_{b}) + \frac{v_{abcd} + v_{abdc}}{\sqrt{1+\delta_{ab}}\sqrt{1+\delta_{cd}}}
	\\ 
	\label{eq:1B_ppRPA}
	{}^1B_{ab,ij}^{\ee,\hh} & = \frac{v_{abij} + v_{abji}}{\sqrt{1+\delta_{ab}}\sqrt{1+\delta_{ij}}}
	\\ 
	\label{eq:1C_ppRPA}
	{}^1C_{ij,kl}^{\hh} & = - \delta_{ik} \delta_{jl} (\e_{i} + \e_{j}) + \frac{v_{ijkl} + v_{ijlk}}{\sqrt{1+\delta_{ij}}\sqrt{1+\delta_{kl}}}
\end{align}
\end{subequations}
with the index restrictions $a \le b$, $c \le d$, $i \le j$, and $k \le l$, while, for the triplet manifold, it yields three identical sets of equations with matrix elements
\begin{subequations}
\begin{align}
	\label{eq:3A_ppRPA}
	{}^3A_{ab,cd}^{\ee} & = \delta_{ac} \delta_{bd} (\e_{a} + \e_{b}) + v_{abcd} - v_{abdc}
	\\ 
	\label{eq:3B_ppRPA}
	{}^3B_{ab,ij}^{\ee,\hh} & = v_{abij} - v_{abji}
	\\ 
	\label{eq:3C_ppRPA}
	{}^3C_{ij,kl}^{\hh} & = - \delta_{ik} \delta_{jl} (\e_{i} + \e_{j}) + v_{ijkl} - v_{ijlk}
\end{align}
\end{subequations}
with the index restrictions $a < b$, $c < d$, $i < j$, and $k < l$.
We refer the interested reader to the work of Yang \textit{et al.} for additional details concerning the spin adaptation of the pp-RPA equations. \cite{Yang_2013}

\bibliography{ehT}

\begin{thebibliography}{98}%
\makeatletter
\providecommand \@ifxundefined [1]{%
 \@ifx{#1\undefined}
}%
\providecommand \@ifnum [1]{%
 \ifnum #1\expandafter \@firstoftwo
 \else \expandafter \@secondoftwo
 \fi
}%
\providecommand \@ifx [1]{%
 \ifx #1\expandafter \@firstoftwo
 \else \expandafter \@secondoftwo
 \fi
}%
\providecommand \natexlab [1]{#1}%
\providecommand \enquote  [1]{``#1''}%
\providecommand \bibnamefont  [1]{#1}%
\providecommand \bibfnamefont [1]{#1}%
\providecommand \citenamefont [1]{#1}%
\providecommand \href@noop [0]{\@secondoftwo}%
\providecommand \href [0]{\begingroup \@sanitize@url \@href}%
\providecommand \@href[1]{\@@startlink{#1}\@@href}%
\providecommand \@@href[1]{\endgroup#1\@@endlink}%
\providecommand \@sanitize@url [0]{\catcode `\\12\catcode `\$12\catcode
  `\&12\catcode `\#12\catcode `\^12\catcode `\_12\catcode `\%12\relax}%
\providecommand \@@startlink[1]{}%
\providecommand \@@endlink[0]{}%
\providecommand \url  [0]{\begingroup\@sanitize@url \@url }%
\providecommand \@url [1]{\endgroup\@href {#1}{\urlprefix }}%
\providecommand \urlprefix  [0]{URL }%
\providecommand \Eprint [0]{\href }%
\providecommand \doibase [0]{http://dx.doi.org/}%
\providecommand \selectlanguage [0]{\@gobble}%
\providecommand \bibinfo  [0]{\@secondoftwo}%
\providecommand \bibfield  [0]{\@secondoftwo}%
\providecommand \translation [1]{[#1]}%
\providecommand \BibitemOpen [0]{}%
\providecommand \bibitemStop [0]{}%
\providecommand \bibitemNoStop [0]{.\EOS\space}%
\providecommand \EOS [0]{\spacefactor3000\relax}%
\providecommand \BibitemShut  [1]{\csname bibitem#1\endcsname}%
\let\auto@bib@innerbib\@empty
\bibitem [{\citenamefont {Csanak}, \citenamefont {Taylor},\ and\ \citenamefont
  {Yaris}(1971)}]{CsanakBook}%
  \BibitemOpen
  \bibfield  {author} {\bibinfo {author} {\bibfnamefont {G.}~\bibnamefont
  {Csanak}}, \bibinfo {author} {\bibfnamefont {H.}~\bibnamefont {Taylor}}, \
  and\ \bibinfo {author} {\bibfnamefont {R.}~\bibnamefont {Yaris}},\ }in\
  \href@noop {} {\emph {\bibinfo {booktitle} {Advances in atomic and molecular
  physics}}},\ Vol.~\bibinfo {volume} {7}\ (\bibinfo  {publisher} {Elsevier},\
  \bibinfo {year} {1971})\ pp.\ \bibinfo {pages} {287--361}\BibitemShut
  {NoStop}%
\bibitem [{\citenamefont {Fetter}\ and\ \citenamefont
  {Waleck}(1971)}]{FetterBook}%
  \BibitemOpen
  \bibfield  {author} {\bibinfo {author} {\bibfnamefont {A.~L.}\ \bibnamefont
  {Fetter}}\ and\ \bibinfo {author} {\bibfnamefont {J.~D.}\ \bibnamefont
  {Waleck}},\ }\href@noop {} {\emph {\bibinfo {title} {Quantum Theory of Many
  Particle Systems}}}\ (\bibinfo  {publisher} {McGraw Hill, San Francisco},\
  \bibinfo {year} {1971})\BibitemShut {NoStop}%
\bibitem [{\citenamefont {Mattuck}(1992)}]{MattuckBook}%
  \BibitemOpen
  \bibfield  {author} {\bibinfo {author} {\bibfnamefont {R.~D.}\ \bibnamefont
  {Mattuck}},\ }\href@noop {} {\emph {\bibinfo {title} {A guide to {Feynman}
  diagrams in the many-body problem}}},\ \bibinfo {edition} {2nd}\ ed.,\ Dover
  books on physics and chemistry\ (\bibinfo  {publisher} {Dover Publications},\
  \bibinfo {address} {New York},\ \bibinfo {year} {1992})\BibitemShut {NoStop}%
\bibitem [{\citenamefont {Martin}, \citenamefont {Reining},\ and\ \citenamefont
  {Ceperley}(2016)}]{MartinBook}%
  \BibitemOpen
  \bibfield  {author} {\bibinfo {author} {\bibfnamefont {R.~M.}\ \bibnamefont
  {Martin}}, \bibinfo {author} {\bibfnamefont {L.}~\bibnamefont {Reining}}, \
  and\ \bibinfo {author} {\bibfnamefont {D.~M.}\ \bibnamefont {Ceperley}},\
  }\href@noop {} {\emph {\bibinfo {title} {Interacting Electrons: Theory and
  Computational Approaches}}}\ (\bibinfo  {publisher} {Cambridge University
  Press},\ \bibinfo {year} {2016})\BibitemShut {NoStop}%
\bibitem [{\citenamefont {Hedin}(1965)}]{Hedin_1965}%
  \BibitemOpen
  \bibfield  {author} {\bibinfo {author} {\bibfnamefont {L.}~\bibnamefont
  {Hedin}},\ }\href {\doibase 10.1103/PhysRev.139.A796} {\bibfield  {journal}
  {\bibinfo  {journal} {Phys. Rev.}\ }\textbf {\bibinfo {volume} {139}},\
  \bibinfo {pages} {A796} (\bibinfo {year} {1965})}\BibitemShut {NoStop}%
\bibitem [{\citenamefont {Aryasetiawan}\ and\ \citenamefont
  {Gunnarsson}(1998)}]{Aryasetiawan_1998}%
  \BibitemOpen
  \bibfield  {author} {\bibinfo {author} {\bibfnamefont {F.}~\bibnamefont
  {Aryasetiawan}}\ and\ \bibinfo {author} {\bibfnamefont {O.}~\bibnamefont
  {Gunnarsson}},\ }\href {\doibase 10.1088/0034-4885/61/3/002} {\bibfield
  {journal} {\bibinfo  {journal} {Rep. Prog. Phys.}\ }\textbf {\bibinfo
  {volume} {61}},\ \bibinfo {pages} {237} (\bibinfo {year} {1998})}\BibitemShut
  {NoStop}%
\bibitem [{\citenamefont {Onida}, \citenamefont {Reining},\ and\ \citenamefont
  {Rubio}(2002)}]{Onida_2002}%
  \BibitemOpen
  \bibfield  {author} {\bibinfo {author} {\bibfnamefont {G.}~\bibnamefont
  {Onida}}, \bibinfo {author} {\bibfnamefont {L.}~\bibnamefont {Reining}}, \
  and\ \bibinfo {author} {\bibfnamefont {A.}~\bibnamefont {Rubio}},\ }\href
  {\doibase 10.1103/RevModPhys.74.601} {\bibfield  {journal} {\bibinfo
  {journal} {Rev. Mod. Phys.}\ }\textbf {\bibinfo {volume} {74}},\ \bibinfo
  {pages} {601} (\bibinfo {year} {2002})}\BibitemShut {NoStop}%
\bibitem [{\citenamefont {Reining}(2017)}]{Reining_2017}%
  \BibitemOpen
  \bibfield  {author} {\bibinfo {author} {\bibfnamefont {L.}~\bibnamefont
  {Reining}},\ }\href {\doibase 10.1002/wcms.1344} {\bibfield  {journal}
  {\bibinfo  {journal} {WIREs Comput. Mol. Sci.}\ }\textbf {\bibinfo {volume}
  {8}},\ \bibinfo {pages} {e1344} (\bibinfo {year} {2017})}\BibitemShut
  {NoStop}%
\bibitem [{\citenamefont {Golze}, \citenamefont {Dvorak},\ and\ \citenamefont
  {Rinke}(2019)}]{Golze_2019}%
  \BibitemOpen
  \bibfield  {author} {\bibinfo {author} {\bibfnamefont {D.}~\bibnamefont
  {Golze}}, \bibinfo {author} {\bibfnamefont {M.}~\bibnamefont {Dvorak}}, \
  and\ \bibinfo {author} {\bibfnamefont {P.}~\bibnamefont {Rinke}},\ }\href
  {\doibase 10.3389/fchem.2019.00377} {\bibfield  {journal} {\bibinfo
  {journal} {Front. Chem.}\ }\textbf {\bibinfo {volume} {7}},\ \bibinfo {pages}
  {377} (\bibinfo {year} {2019})}\BibitemShut {NoStop}%
\bibitem [{\citenamefont {Gell-Mann}\ and\ \citenamefont
  {Brueckner}(1957)}]{Gell-Mann_1957}%
  \BibitemOpen
  \bibfield  {author} {\bibinfo {author} {\bibfnamefont {M.}~\bibnamefont
  {Gell-Mann}}\ and\ \bibinfo {author} {\bibfnamefont {K.~A.}\ \bibnamefont
  {Brueckner}},\ }\href {\doibase 10.1103/PhysRev.106.364} {\bibfield
  {journal} {\bibinfo  {journal} {Phys. Rev.}\ }\textbf {\bibinfo {volume}
  {106}},\ \bibinfo {pages} {364} (\bibinfo {year} {1957})}\BibitemShut
  {NoStop}%
\bibitem [{\citenamefont {Baym}\ and\ \citenamefont
  {Kadanoff}(1961)}]{Baym_1961}%
  \BibitemOpen
  \bibfield  {author} {\bibinfo {author} {\bibfnamefont {G.}~\bibnamefont
  {Baym}}\ and\ \bibinfo {author} {\bibfnamefont {L.~P.}\ \bibnamefont
  {Kadanoff}},\ }\href {\doibase 10.1103/PhysRev.124.287} {\bibfield  {journal}
  {\bibinfo  {journal} {Phys. Rev.}\ }\textbf {\bibinfo {volume} {124}},\
  \bibinfo {pages} {287} (\bibinfo {year} {1961})}\BibitemShut {NoStop}%
\bibitem [{\citenamefont {Baym}(1962)}]{Baym_1962}%
  \BibitemOpen
  \bibfield  {author} {\bibinfo {author} {\bibfnamefont {G.}~\bibnamefont
  {Baym}},\ }\href {\doibase 10.1103/PhysRev.127.1391} {\bibfield  {journal}
  {\bibinfo  {journal} {Phys. Rev.}\ }\textbf {\bibinfo {volume} {127}},\
  \bibinfo {pages} {1391} (\bibinfo {year} {1962})}\BibitemShut {NoStop}%
\bibitem [{\citenamefont
  {Danielewicz}(1984{\natexlab{a}})}]{Danielewicz_1984a}%
  \BibitemOpen
  \bibfield  {author} {\bibinfo {author} {\bibfnamefont {P.}~\bibnamefont
  {Danielewicz}},\ }\href {\doibase
  https://doi.org/10.1016/0003-4916(84)90092-7} {\bibfield  {journal} {\bibinfo
   {journal} {Ann. Phys.}\ }\textbf {\bibinfo {volume} {152}},\ \bibinfo
  {pages} {239} (\bibinfo {year} {1984}{\natexlab{a}})}\BibitemShut {NoStop}%
\bibitem [{\citenamefont
  {Danielewicz}(1984{\natexlab{b}})}]{Danielewicz_1984b}%
  \BibitemOpen
  \bibfield  {author} {\bibinfo {author} {\bibfnamefont {P.}~\bibnamefont
  {Danielewicz}},\ }\href {\doibase
  https://doi.org/10.1016/0003-4916(84)90093-9} {\bibfield  {journal} {\bibinfo
   {journal} {Ann. Phys.}\ }\textbf {\bibinfo {volume} {152}},\ \bibinfo
  {pages} {305} (\bibinfo {year} {1984}{\natexlab{b}})}\BibitemShut {NoStop}%
\bibitem [{\citenamefont {Bohm}\ and\ \citenamefont {Pines}(1951)}]{Bohm_1951}%
  \BibitemOpen
  \bibfield  {author} {\bibinfo {author} {\bibfnamefont {D.}~\bibnamefont
  {Bohm}}\ and\ \bibinfo {author} {\bibfnamefont {D.}~\bibnamefont {Pines}},\
  }\href {\doibase 10.1103/PhysRev.82.625} {\bibfield  {journal} {\bibinfo
  {journal} {Phys. Rev.}\ }\textbf {\bibinfo {volume} {82}},\ \bibinfo {pages}
  {625} (\bibinfo {year} {1951})}\BibitemShut {NoStop}%
\bibitem [{\citenamefont {Pines}\ and\ \citenamefont
  {Bohm}(1952)}]{Pines_1952}%
  \BibitemOpen
  \bibfield  {author} {\bibinfo {author} {\bibfnamefont {D.}~\bibnamefont
  {Pines}}\ and\ \bibinfo {author} {\bibfnamefont {D.}~\bibnamefont {Bohm}},\
  }\href {\doibase 10.1103/PhysRev.85.338} {\bibfield  {journal} {\bibinfo
  {journal} {Phys. Rev.}\ }\textbf {\bibinfo {volume} {85}},\ \bibinfo {pages}
  {338} (\bibinfo {year} {1952})}\BibitemShut {NoStop}%
\bibitem [{\citenamefont {Bohm}\ and\ \citenamefont {Pines}(1953)}]{Bohm_1953}%
  \BibitemOpen
  \bibfield  {author} {\bibinfo {author} {\bibfnamefont {D.}~\bibnamefont
  {Bohm}}\ and\ \bibinfo {author} {\bibfnamefont {D.}~\bibnamefont {Pines}},\
  }\href {\doibase 10.1103/PhysRev.92.609} {\bibfield  {journal} {\bibinfo
  {journal} {Phys. Rev.}\ }\textbf {\bibinfo {volume} {92}},\ \bibinfo {pages}
  {609} (\bibinfo {year} {1953})}\BibitemShut {NoStop}%
\bibitem [{\citenamefont {Nozi\`eres}\ and\ \citenamefont
  {Pines}(1958)}]{Nozieres_1958}%
  \BibitemOpen
  \bibfield  {author} {\bibinfo {author} {\bibfnamefont {P.}~\bibnamefont
  {Nozi\`eres}}\ and\ \bibinfo {author} {\bibfnamefont {D.}~\bibnamefont
  {Pines}},\ }\href {\doibase 10.1103/PhysRev.111.442} {\bibfield  {journal}
  {\bibinfo  {journal} {Phys. Rev.}\ }\textbf {\bibinfo {volume} {111}},\
  \bibinfo {pages} {442} (\bibinfo {year} {1958})}\BibitemShut {NoStop}%
\bibitem [{\citenamefont {Romaniello}, \citenamefont {Bechstedt},\ and\
  \citenamefont {Reining}(2012)}]{Romaniello_2012}%
  \BibitemOpen
  \bibfield  {author} {\bibinfo {author} {\bibfnamefont {P.}~\bibnamefont
  {Romaniello}}, \bibinfo {author} {\bibfnamefont {F.}~\bibnamefont
  {Bechstedt}}, \ and\ \bibinfo {author} {\bibfnamefont {L.}~\bibnamefont
  {Reining}},\ }\href {\doibase 10.1103/PhysRevB.85.155131} {\bibfield
  {journal} {\bibinfo  {journal} {Phys. Rev. B}\ }\textbf {\bibinfo {volume}
  {85}},\ \bibinfo {pages} {155131} (\bibinfo {year} {2012})}\BibitemShut
  {NoStop}%
\bibitem [{\citenamefont {Starke}\ and\ \citenamefont
  {Kresse}(2012)}]{Starke_2012}%
  \BibitemOpen
  \bibfield  {author} {\bibinfo {author} {\bibfnamefont {R.}~\bibnamefont
  {Starke}}\ and\ \bibinfo {author} {\bibfnamefont {G.}~\bibnamefont
  {Kresse}},\ }\href {\doibase 10.1103/PhysRevB.85.075119} {\bibfield
  {journal} {\bibinfo  {journal} {Phys. Rev. B}\ }\textbf {\bibinfo {volume}
  {85}},\ \bibinfo {pages} {075119} (\bibinfo {year} {2012})}\BibitemShut
  {NoStop}%
\bibitem [{\citenamefont {Maggio}\ and\ \citenamefont
  {Kresse}(2017)}]{Maggio_2017b}%
  \BibitemOpen
  \bibfield  {author} {\bibinfo {author} {\bibfnamefont {E.}~\bibnamefont
  {Maggio}}\ and\ \bibinfo {author} {\bibfnamefont {G.}~\bibnamefont
  {Kresse}},\ }\href {\doibase 10.1021/acs.jctc.7b00586} {\bibfield  {journal}
  {\bibinfo  {journal} {J. Chem. Theory Comput.}\ }\textbf {\bibinfo {volume}
  {13}},\ \bibinfo {pages} {4765} (\bibinfo {year} {2017})}\BibitemShut
  {NoStop}%
\bibitem [{\citenamefont {Del~Sole}, \citenamefont {Reining},\ and\
  \citenamefont {Godby}(1994)}]{DelSol_1994}%
  \BibitemOpen
  \bibfield  {author} {\bibinfo {author} {\bibfnamefont {R.}~\bibnamefont
  {Del~Sole}}, \bibinfo {author} {\bibfnamefont {L.}~\bibnamefont {Reining}}, \
  and\ \bibinfo {author} {\bibfnamefont {R.~W.}\ \bibnamefont {Godby}},\ }\href
  {\doibase 10.1103/PhysRevB.49.8024} {\bibfield  {journal} {\bibinfo
  {journal} {Phys. Rev. B}\ }\textbf {\bibinfo {volume} {49}},\ \bibinfo
  {pages} {8024} (\bibinfo {year} {1994})}\BibitemShut {NoStop}%
\bibitem [{\citenamefont {Shirley}(1996)}]{Shirley_1996}%
  \BibitemOpen
  \bibfield  {author} {\bibinfo {author} {\bibfnamefont {E.~L.}\ \bibnamefont
  {Shirley}},\ }\href {\doibase 10.1103/PhysRevB.54.7758} {\bibfield  {journal}
  {\bibinfo  {journal} {Phys. Rev. B}\ }\textbf {\bibinfo {volume} {54}},\
  \bibinfo {pages} {7758} (\bibinfo {year} {1996})}\BibitemShut {NoStop}%
\bibitem [{\citenamefont {Schindlmayr}\ and\ \citenamefont
  {Godby}(1998)}]{Schindlmayr_1998}%
  \BibitemOpen
  \bibfield  {author} {\bibinfo {author} {\bibfnamefont {A.}~\bibnamefont
  {Schindlmayr}}\ and\ \bibinfo {author} {\bibfnamefont {R.~W.}\ \bibnamefont
  {Godby}},\ }\href {\doibase 10.1103/PhysRevLett.80.1702} {\bibfield
  {journal} {\bibinfo  {journal} {Phys. Rev. Lett.}\ }\textbf {\bibinfo
  {volume} {80}},\ \bibinfo {pages} {1702} (\bibinfo {year}
  {1998})}\BibitemShut {NoStop}%
\bibitem [{\citenamefont {Morris}\ \emph {et~al.}(2007)\citenamefont {Morris},
  \citenamefont {Stankovski}, \citenamefont {Delaney}, \citenamefont {Rinke},
  \citenamefont {Garc\'{\i}a-Gonz\'alez},\ and\ \citenamefont
  {Godby}}]{Morris_2007}%
  \BibitemOpen
  \bibfield  {author} {\bibinfo {author} {\bibfnamefont {A.~J.}\ \bibnamefont
  {Morris}}, \bibinfo {author} {\bibfnamefont {M.}~\bibnamefont {Stankovski}},
  \bibinfo {author} {\bibfnamefont {K.~T.}\ \bibnamefont {Delaney}}, \bibinfo
  {author} {\bibfnamefont {P.}~\bibnamefont {Rinke}}, \bibinfo {author}
  {\bibfnamefont {P.}~\bibnamefont {Garc\'{\i}a-Gonz\'alez}}, \ and\ \bibinfo
  {author} {\bibfnamefont {R.~W.}\ \bibnamefont {Godby}},\ }\href {\doibase
  10.1103/PhysRevB.76.155106} {\bibfield  {journal} {\bibinfo  {journal} {Phys.
  Rev. B}\ }\textbf {\bibinfo {volume} {76}},\ \bibinfo {pages} {155106}
  (\bibinfo {year} {2007})}\BibitemShut {NoStop}%
\bibitem [{\citenamefont {Shishkin}, \citenamefont {Marsman},\ and\
  \citenamefont {Kresse}(2007)}]{Shishkin_2007b}%
  \BibitemOpen
  \bibfield  {author} {\bibinfo {author} {\bibfnamefont {M.}~\bibnamefont
  {Shishkin}}, \bibinfo {author} {\bibfnamefont {M.}~\bibnamefont {Marsman}}, \
  and\ \bibinfo {author} {\bibfnamefont {G.}~\bibnamefont {Kresse}},\ }\href
  {\doibase 10.1103/PhysRevLett.99.246403} {\bibfield  {journal} {\bibinfo
  {journal} {Phys. Rev. Lett.}\ }\textbf {\bibinfo {volume} {99}},\ \bibinfo
  {pages} {246403} (\bibinfo {year} {2007})}\BibitemShut {NoStop}%
\bibitem [{\citenamefont {Romaniello}, \citenamefont {Guyot},\ and\
  \citenamefont {Reining}(2009)}]{Romaniello_2009a}%
  \BibitemOpen
  \bibfield  {author} {\bibinfo {author} {\bibfnamefont {P.}~\bibnamefont
  {Romaniello}}, \bibinfo {author} {\bibfnamefont {S.}~\bibnamefont {Guyot}}, \
  and\ \bibinfo {author} {\bibfnamefont {L.}~\bibnamefont {Reining}},\ }\href
  {\doibase 10.1063/1.3249965} {\bibfield  {journal} {\bibinfo  {journal} {J.
  Chem. Phys.}\ }\textbf {\bibinfo {volume} {131}},\ \bibinfo {pages} {154111}
  (\bibinfo {year} {2009})}\BibitemShut {NoStop}%
\bibitem [{\citenamefont {Gr\"uneis}\ \emph {et~al.}(2014)\citenamefont
  {Gr\"uneis}, \citenamefont {Kresse}, \citenamefont {Hinuma},\ and\
  \citenamefont {Oba}}]{Gruneis_2014}%
  \BibitemOpen
  \bibfield  {author} {\bibinfo {author} {\bibfnamefont {A.}~\bibnamefont
  {Gr\"uneis}}, \bibinfo {author} {\bibfnamefont {G.}~\bibnamefont {Kresse}},
  \bibinfo {author} {\bibfnamefont {Y.}~\bibnamefont {Hinuma}}, \ and\ \bibinfo
  {author} {\bibfnamefont {F.}~\bibnamefont {Oba}},\ }\href {\doibase
  10.1103/PhysRevLett.112.096401} {\bibfield  {journal} {\bibinfo  {journal}
  {Phys. Rev. Lett.}\ }\textbf {\bibinfo {volume} {112}},\ \bibinfo {pages}
  {096401} (\bibinfo {year} {2014})}\BibitemShut {NoStop}%
\bibitem [{\citenamefont {Chen}\ and\ \citenamefont
  {Pasquarello}(2015)}]{Chen_2015}%
  \BibitemOpen
  \bibfield  {author} {\bibinfo {author} {\bibfnamefont {W.}~\bibnamefont
  {Chen}}\ and\ \bibinfo {author} {\bibfnamefont {A.}~\bibnamefont
  {Pasquarello}},\ }\href {\doibase 10.1103/PhysRevB.92.041115} {\bibfield
  {journal} {\bibinfo  {journal} {Phys. Rev. B}\ }\textbf {\bibinfo {volume}
  {92}},\ \bibinfo {pages} {041115} (\bibinfo {year} {2015})}\BibitemShut
  {NoStop}%
\bibitem [{\citenamefont {Ren}\ \emph {et~al.}(2015)\citenamefont {Ren},
  \citenamefont {Marom}, \citenamefont {Caruso}, \citenamefont {Scheffler},\
  and\ \citenamefont {Rinke}}]{Ren_2015}%
  \BibitemOpen
  \bibfield  {author} {\bibinfo {author} {\bibfnamefont {X.}~\bibnamefont
  {Ren}}, \bibinfo {author} {\bibfnamefont {N.}~\bibnamefont {Marom}}, \bibinfo
  {author} {\bibfnamefont {F.}~\bibnamefont {Caruso}}, \bibinfo {author}
  {\bibfnamefont {M.}~\bibnamefont {Scheffler}}, \ and\ \bibinfo {author}
  {\bibfnamefont {P.}~\bibnamefont {Rinke}},\ }\href {\doibase
  10.1103/PhysRevB.92.081104} {\bibfield  {journal} {\bibinfo  {journal} {Phys.
  Rev. B}\ }\textbf {\bibinfo {volume} {92}},\ \bibinfo {pages} {081104}
  (\bibinfo {year} {2015})}\BibitemShut {NoStop}%
\bibitem [{\citenamefont {Caruso}\ \emph {et~al.}(2016)\citenamefont {Caruso},
  \citenamefont {Dauth}, \citenamefont {{van Setten}},\ and\ \citenamefont
  {Rinke}}]{Caruso_2016}%
  \BibitemOpen
  \bibfield  {author} {\bibinfo {author} {\bibfnamefont {F.}~\bibnamefont
  {Caruso}}, \bibinfo {author} {\bibfnamefont {M.}~\bibnamefont {Dauth}},
  \bibinfo {author} {\bibfnamefont {M.~J.}\ \bibnamefont {{van Setten}}}, \
  and\ \bibinfo {author} {\bibfnamefont {P.}~\bibnamefont {Rinke}},\ }\href
  {\doibase 10.1021/acs.jctc.6b00774} {\bibfield  {journal} {\bibinfo
  {journal} {J. Chem. Theory Comput.}\ }\textbf {\bibinfo {volume} {12}},\
  \bibinfo {pages} {5076} (\bibinfo {year} {2016})}\BibitemShut {NoStop}%
\bibitem [{\citenamefont {Hung}\ \emph {et~al.}(2017)\citenamefont {Hung},
  \citenamefont {Bruneval}, \citenamefont {Baishya},\ and\ \citenamefont
  {{\"O}{\u g}{\"u}t}}]{Hung_2017}%
  \BibitemOpen
  \bibfield  {author} {\bibinfo {author} {\bibfnamefont {L.}~\bibnamefont
  {Hung}}, \bibinfo {author} {\bibfnamefont {F.}~\bibnamefont {Bruneval}},
  \bibinfo {author} {\bibfnamefont {K.}~\bibnamefont {Baishya}}, \ and\
  \bibinfo {author} {\bibfnamefont {S.}~\bibnamefont {{\"O}{\u g}{\"u}t}},\
  }\href {\doibase 10.1021/acs.jctc.7b00123} {\bibfield  {journal} {\bibinfo
  {journal} {J. Chem. Theory Comput.}\ }\textbf {\bibinfo {volume} {13}},\
  \bibinfo {pages} {2135} (\bibinfo {year} {2017})}\BibitemShut {NoStop}%
\bibitem [{\citenamefont {Cunningham}\ \emph {et~al.}(2018)\citenamefont
  {Cunningham}, \citenamefont {Gr\"uning}, \citenamefont {Azarhoosh},
  \citenamefont {Pashov},\ and\ \citenamefont {van
  Schilfgaarde}}]{Cunningham_2018}%
  \BibitemOpen
  \bibfield  {author} {\bibinfo {author} {\bibfnamefont {B.}~\bibnamefont
  {Cunningham}}, \bibinfo {author} {\bibfnamefont {M.}~\bibnamefont
  {Gr\"uning}}, \bibinfo {author} {\bibfnamefont {P.}~\bibnamefont
  {Azarhoosh}}, \bibinfo {author} {\bibfnamefont {D.}~\bibnamefont {Pashov}}, \
  and\ \bibinfo {author} {\bibfnamefont {M.}~\bibnamefont {van Schilfgaarde}},\
  }\href {\doibase 10.1103/PhysRevMaterials.2.034603} {\bibfield  {journal}
  {\bibinfo  {journal} {Phys. Rev. Mater.}\ }\textbf {\bibinfo {volume} {2}},\
  \bibinfo {pages} {034603} (\bibinfo {year} {2018})}\BibitemShut {NoStop}%
\bibitem [{\citenamefont {Vl{\v c}ek}(2019)}]{Vlcek_2019}%
  \BibitemOpen
  \bibfield  {author} {\bibinfo {author} {\bibfnamefont {V.}~\bibnamefont
  {Vl{\v c}ek}},\ }\href {\doibase https://doi.org/10.1021/acs.jctc.9b00317}
  {\bibfield  {journal} {\bibinfo  {journal} {J. Chem. Theory Comput.}\
  }\textbf {\bibinfo {volume} {15}},\ \bibinfo {pages} {6254} (\bibinfo {year}
  {2019})}\BibitemShut {NoStop}%
\bibitem [{\citenamefont {Lewis}\ and\ \citenamefont
  {Berkelbach}(2019)}]{Lewis_2019a}%
  \BibitemOpen
  \bibfield  {author} {\bibinfo {author} {\bibfnamefont {A.~M.}\ \bibnamefont
  {Lewis}}\ and\ \bibinfo {author} {\bibfnamefont {T.~C.}\ \bibnamefont
  {Berkelbach}},\ }\href {\doibase 10.1021/acs.jctc.8b00995} {\bibfield
  {journal} {\bibinfo  {journal} {J. Chem. Theory Comput.}\ }\textbf {\bibinfo
  {volume} {15}},\ \bibinfo {pages} {2925} (\bibinfo {year}
  {2019})}\BibitemShut {NoStop}%
\bibitem [{\citenamefont {Pavlyukh}, \citenamefont {Stefanucci},\ and\
  \citenamefont {van Leeuwen}(2020)}]{Pavlyukh_2020}%
  \BibitemOpen
  \bibfield  {author} {\bibinfo {author} {\bibfnamefont {Y.}~\bibnamefont
  {Pavlyukh}}, \bibinfo {author} {\bibfnamefont {G.}~\bibnamefont
  {Stefanucci}}, \ and\ \bibinfo {author} {\bibfnamefont {R.}~\bibnamefont {van
  Leeuwen}},\ }\href {\doibase 10.1103/PhysRevB.102.045121} {\bibfield
  {journal} {\bibinfo  {journal} {Phys. Rev. B}\ }\textbf {\bibinfo {volume}
  {102}},\ \bibinfo {pages} {045121} (\bibinfo {year} {2020})}\BibitemShut
  {NoStop}%
\bibitem [{\citenamefont {Wang}, \citenamefont {Rinke},\ and\ \citenamefont
  {Ren}(2021)}]{Wang_2021}%
  \BibitemOpen
  \bibfield  {author} {\bibinfo {author} {\bibfnamefont {Y.}~\bibnamefont
  {Wang}}, \bibinfo {author} {\bibfnamefont {P.}~\bibnamefont {Rinke}}, \ and\
  \bibinfo {author} {\bibfnamefont {X.}~\bibnamefont {Ren}},\ }\href {\doibase
  10.1021/acs.jctc.1c00488} {\bibfield  {journal} {\bibinfo  {journal} {J.
  Chem. Theory Comput.}\ }\textbf {\bibinfo {volume} {17}},\ \bibinfo {pages}
  {5140} (\bibinfo {year} {2021})}\BibitemShut {NoStop}%
\bibitem [{\citenamefont {Bruneval}, \citenamefont {Dattani},\ and\
  \citenamefont {van Setten}(2021)}]{Bruneval_2021}%
  \BibitemOpen
  \bibfield  {author} {\bibinfo {author} {\bibfnamefont {F.}~\bibnamefont
  {Bruneval}}, \bibinfo {author} {\bibfnamefont {N.}~\bibnamefont {Dattani}}, \
  and\ \bibinfo {author} {\bibfnamefont {M.~J.}\ \bibnamefont {van Setten}},\
  }\href {\doibase 10.3389/fchem.2021.749779} {\bibfield  {journal} {\bibinfo
  {journal} {Front. Chem.}\ }\textbf {\bibinfo {volume} {9}},\ \bibinfo {pages}
  {749779} (\bibinfo {year} {2021})}\BibitemShut {NoStop}%
\bibitem [{\citenamefont {Mejuto-Zaera}\ and\ \citenamefont
  {Vl\ifmmode~\check{c}\else \v{c}\fi{}ek}(2022)}]{Mejuto-Zaera_2022}%
  \BibitemOpen
  \bibfield  {author} {\bibinfo {author} {\bibfnamefont {C.}~\bibnamefont
  {Mejuto-Zaera}}\ and\ \bibinfo {author} {\bibfnamefont {V.~c.~v.}\
  \bibnamefont {Vl\ifmmode~\check{c}\else \v{c}\fi{}ek}},\ }\href {\doibase
  10.1103/PhysRevB.106.165129} {\bibfield  {journal} {\bibinfo  {journal}
  {Phys. Rev. B}\ }\textbf {\bibinfo {volume} {106}},\ \bibinfo {pages}
  {165129} (\bibinfo {year} {2022})}\BibitemShut {NoStop}%
\bibitem [{\citenamefont {Wang}\ and\ \citenamefont {Ren}(2022)}]{Wang_2022}%
  \BibitemOpen
  \bibfield  {author} {\bibinfo {author} {\bibfnamefont {Y.}~\bibnamefont
  {Wang}}\ and\ \bibinfo {author} {\bibfnamefont {X.}~\bibnamefont {Ren}},\
  }\href {\doibase 10.1063/5.0122425} {\bibfield  {journal} {\bibinfo
  {journal} {J. Chem. Theory Comput.}\ }\textbf {\bibinfo {volume} {157}},\
  \bibinfo {pages} {214115} (\bibinfo {year} {2022})}\BibitemShut {NoStop}%
\bibitem [{\citenamefont {F{\"o}rster}\ and\ \citenamefont
  {Visscher}(2022)}]{Forster_2022b}%
  \BibitemOpen
  \bibfield  {author} {\bibinfo {author} {\bibfnamefont {A.}~\bibnamefont
  {F{\"o}rster}}\ and\ \bibinfo {author} {\bibfnamefont {L.}~\bibnamefont
  {Visscher}},\ }\href {\doibase 10.1103/PhysRevB.105.125121} {\bibfield
  {journal} {\bibinfo  {journal} {Phys. Rev. B}\ }\textbf {\bibinfo {volume}
  {105}},\ \bibinfo {pages} {125121} (\bibinfo {year} {2022})}\BibitemShut
  {NoStop}%
\bibitem [{\citenamefont {Dadkhah}\ \emph {et~al.}(2023)\citenamefont
  {Dadkhah}, \citenamefont {Lambrecht}, \citenamefont {Pashov},\ and\
  \citenamefont {van Schilfgaarde}}]{Dadkhah_2023}%
  \BibitemOpen
  \bibfield  {author} {\bibinfo {author} {\bibfnamefont {N.}~\bibnamefont
  {Dadkhah}}, \bibinfo {author} {\bibfnamefont {W.~R.~L.}\ \bibnamefont
  {Lambrecht}}, \bibinfo {author} {\bibfnamefont {D.}~\bibnamefont {Pashov}}, \
  and\ \bibinfo {author} {\bibfnamefont {M.}~\bibnamefont {van Schilfgaarde}},\
  }\href {\doibase 10.1103/PhysRevB.107.165201} {\bibfield  {journal} {\bibinfo
   {journal} {Phys. Rev. B}\ }\textbf {\bibinfo {volume} {107}},\ \bibinfo
  {pages} {165201} (\bibinfo {year} {2023})}\BibitemShut {NoStop}%
\bibitem [{\citenamefont {Grzeszczyk}\ \emph {et~al.}(2023)\citenamefont
  {Grzeszczyk}, \citenamefont {Acharya}, \citenamefont {Pashov}, \citenamefont
  {Chen}, \citenamefont {Vaklinova}, \citenamefont {van Schilfgaarde},
  \citenamefont {Watanabe}, \citenamefont {Taniguchi}, \citenamefont
  {Novoselov}, \citenamefont {Katsnelson},\ and\ \citenamefont
  {Koperski}}]{Grzeszczyk_2023}%
  \BibitemOpen
  \bibfield  {author} {\bibinfo {author} {\bibfnamefont {M.}~\bibnamefont
  {Grzeszczyk}}, \bibinfo {author} {\bibfnamefont {S.}~\bibnamefont {Acharya}},
  \bibinfo {author} {\bibfnamefont {D.}~\bibnamefont {Pashov}}, \bibinfo
  {author} {\bibfnamefont {Z.}~\bibnamefont {Chen}}, \bibinfo {author}
  {\bibfnamefont {K.}~\bibnamefont {Vaklinova}}, \bibinfo {author}
  {\bibfnamefont {M.}~\bibnamefont {van Schilfgaarde}}, \bibinfo {author}
  {\bibfnamefont {K.}~\bibnamefont {Watanabe}}, \bibinfo {author}
  {\bibfnamefont {T.}~\bibnamefont {Taniguchi}}, \bibinfo {author}
  {\bibfnamefont {K.~S.}\ \bibnamefont {Novoselov}}, \bibinfo {author}
  {\bibfnamefont {M.~I.}\ \bibnamefont {Katsnelson}}, \ and\ \bibinfo {author}
  {\bibfnamefont {M.}~\bibnamefont {Koperski}},\ }\href {\doibase
  https://doi.org/10.1002/adma.202209513} {\bibfield  {journal} {\bibinfo
  {journal} {Adv. Mater.}\ }\textbf {\bibinfo {volume} {35}},\ \bibinfo {pages}
  {2209513} (\bibinfo {year} {2023})}\BibitemShut {NoStop}%
\bibitem [{\citenamefont {Strinati}(1988)}]{Strinati_1988}%
  \BibitemOpen
  \bibfield  {author} {\bibinfo {author} {\bibfnamefont {G.}~\bibnamefont
  {Strinati}},\ }\href {\doibase 10.1007/BF02725962} {\bibfield  {journal}
  {\bibinfo  {journal} {Riv. Nuovo Cimento}\ }\textbf {\bibinfo {volume}
  {11}},\ \bibinfo {pages} {1} (\bibinfo {year} {1988})}\BibitemShut {NoStop}%
\bibitem [{\citenamefont {Ring}\ and\ \citenamefont
  {Schuck}(2004)}]{Schuck_Book}%
  \BibitemOpen
  \bibfield  {author} {\bibinfo {author} {\bibfnamefont {P.}~\bibnamefont
  {Ring}}\ and\ \bibinfo {author} {\bibfnamefont {P.}~\bibnamefont {Schuck}},\
  }\href@noop {} {\emph {\bibinfo {title} {The Nuclear Many-Body Problem}}}\
  (\bibinfo  {publisher} {Springer},\ \bibinfo {year} {2004})\BibitemShut
  {NoStop}%
\bibitem [{\citenamefont {Barbieri}, \citenamefont {Van~Neck},\ and\
  \citenamefont {Dickhoff}(2007)}]{Barbieri_2007}%
  \BibitemOpen
  \bibfield  {author} {\bibinfo {author} {\bibfnamefont {C.}~\bibnamefont
  {Barbieri}}, \bibinfo {author} {\bibfnamefont {D.}~\bibnamefont {Van~Neck}},
  \ and\ \bibinfo {author} {\bibfnamefont {W.~H.}\ \bibnamefont {Dickhoff}},\
  }\href {\doibase 10.1103/PhysRevA.76.052503} {\bibfield  {journal} {\bibinfo
  {journal} {Phys. Rev. A}\ }\textbf {\bibinfo {volume} {76}},\ \bibinfo
  {pages} {052503} (\bibinfo {year} {2007})}\BibitemShut {NoStop}%
\bibitem [{\citenamefont {Dickhoff}\ and\ \citenamefont
  {Neck}(2008)}]{Dickhoff_2008}%
  \BibitemOpen
  \bibfield  {author} {\bibinfo {author} {\bibfnamefont {W.~H.}\ \bibnamefont
  {Dickhoff}}\ and\ \bibinfo {author} {\bibfnamefont {D.~V.}\ \bibnamefont
  {Neck}},\ }\href {\doibase 10.1142/6821} {\emph {\bibinfo {title} {Many-Body
  Theory Exposed!}}}\ (\bibinfo  {publisher} {{WORLD} {SCIENTIFIC}},\ \bibinfo
  {year} {2008})\BibitemShut {NoStop}%
\bibitem [{\citenamefont {Zhang}, \citenamefont {Su},\ and\ \citenamefont
  {Yang}(2017)}]{Zhang_2017}%
  \BibitemOpen
  \bibfield  {author} {\bibinfo {author} {\bibfnamefont {D.}~\bibnamefont
  {Zhang}}, \bibinfo {author} {\bibfnamefont {N.~Q.}\ \bibnamefont {Su}}, \
  and\ \bibinfo {author} {\bibfnamefont {W.}~\bibnamefont {Yang}},\ }\href
  {\doibase 10.1021/acs.jpclett.7b01275} {\bibfield  {journal} {\bibinfo
  {journal} {J. Phys. Chem. Lett.}\ }\textbf {\bibinfo {volume} {8}},\ \bibinfo
  {pages} {3223} (\bibinfo {year} {2017})}\BibitemShut {NoStop}%
\bibitem [{\citenamefont {Li}, \citenamefont {Chen},\ and\ \citenamefont
  {Yang}(2021)}]{Li_2021b}%
  \BibitemOpen
  \bibfield  {author} {\bibinfo {author} {\bibfnamefont {J.}~\bibnamefont
  {Li}}, \bibinfo {author} {\bibfnamefont {Z.}~\bibnamefont {Chen}}, \ and\
  \bibinfo {author} {\bibfnamefont {W.}~\bibnamefont {Yang}},\ }\href {\doibase
  10.1021/acs.jpclett.1c01723} {\bibfield  {journal} {\bibinfo  {journal} {J.
  Phys. Chem. Lett.}\ }\textbf {\bibinfo {volume} {12}},\ \bibinfo {pages}
  {6203} (\bibinfo {year} {2021})}\BibitemShut {NoStop}%
\bibitem [{\citenamefont {Li}\ \emph {et~al.}(2023)\citenamefont {Li},
  \citenamefont {Yu}, \citenamefont {Chen},\ and\ \citenamefont
  {Yang}}]{Li_2023}%
  \BibitemOpen
  \bibfield  {author} {\bibinfo {author} {\bibfnamefont {J.}~\bibnamefont
  {Li}}, \bibinfo {author} {\bibfnamefont {J.}~\bibnamefont {Yu}}, \bibinfo
  {author} {\bibfnamefont {Z.}~\bibnamefont {Chen}}, \ and\ \bibinfo {author}
  {\bibfnamefont {W.}~\bibnamefont {Yang}},\ }\href {\doibase
  10.1021/acs.jpca.3c02834} {\bibfield  {journal} {\bibinfo  {journal} {J.
  Phys. Chem. A}\ }\textbf {\bibinfo {volume} {127}},\ \bibinfo {pages} {7811}
  (\bibinfo {year} {2023})}\BibitemShut {NoStop}%
\bibitem [{\citenamefont {Maggio}\ and\ \citenamefont
  {Kresse}(2018)}]{Maggio_2018}%
  \BibitemOpen
  \bibfield  {author} {\bibinfo {author} {\bibfnamefont {E.}~\bibnamefont
  {Maggio}}\ and\ \bibinfo {author} {\bibfnamefont {G.}~\bibnamefont
  {Kresse}},\ }\href {\doibase 10.1021/acs.jctc.8b00129} {\bibfield  {journal}
  {\bibinfo  {journal} {J. Chem. Theory Comput.}\ }\textbf {\bibinfo {volume}
  {14}},\ \bibinfo {pages} {1821} (\bibinfo {year} {2018})}\BibitemShut
  {NoStop}%
\bibitem [{\citenamefont {Strinati}, \citenamefont {Mattausch},\ and\
  \citenamefont {Hanke}(1980)}]{Strinati_1980}%
  \BibitemOpen
  \bibfield  {author} {\bibinfo {author} {\bibfnamefont {G.}~\bibnamefont
  {Strinati}}, \bibinfo {author} {\bibfnamefont {H.~J.}\ \bibnamefont
  {Mattausch}}, \ and\ \bibinfo {author} {\bibfnamefont {W.}~\bibnamefont
  {Hanke}},\ }\href {\doibase 10.1103/PhysRevLett.45.290} {\bibfield  {journal}
  {\bibinfo  {journal} {Phys. Rev. Lett.}\ }\textbf {\bibinfo {volume} {45}},\
  \bibinfo {pages} {290} (\bibinfo {year} {1980})}\BibitemShut {NoStop}%
\bibitem [{\citenamefont {Hybertsen}\ and\ \citenamefont
  {Louie}(1985)}]{Hybertsen_1985a}%
  \BibitemOpen
  \bibfield  {author} {\bibinfo {author} {\bibfnamefont {M.~S.}\ \bibnamefont
  {Hybertsen}}\ and\ \bibinfo {author} {\bibfnamefont {S.~G.}\ \bibnamefont
  {Louie}},\ }\href {\doibase 10.1103/PhysRevLett.55.1418} {\bibfield
  {journal} {\bibinfo  {journal} {Phys. Rev. Lett.}\ }\textbf {\bibinfo
  {volume} {55}},\ \bibinfo {pages} {1418} (\bibinfo {year}
  {1985})}\BibitemShut {NoStop}%
\bibitem [{\citenamefont {Godby}, \citenamefont {Schl\"uter},\ and\
  \citenamefont {Sham}(1988)}]{Godby_1988}%
  \BibitemOpen
  \bibfield  {author} {\bibinfo {author} {\bibfnamefont {R.~W.}\ \bibnamefont
  {Godby}}, \bibinfo {author} {\bibfnamefont {M.}~\bibnamefont {Schl\"uter}}, \
  and\ \bibinfo {author} {\bibfnamefont {L.~J.}\ \bibnamefont {Sham}},\ }\href
  {\doibase 10.1103/PhysRevB.37.10159} {\bibfield  {journal} {\bibinfo
  {journal} {Phys. Rev. B}\ }\textbf {\bibinfo {volume} {37}},\ \bibinfo
  {pages} {10159} (\bibinfo {year} {1988})}\BibitemShut {NoStop}%
\bibitem [{\citenamefont {von~der Linden}\ and\ \citenamefont
  {Horsch}(1988)}]{Linden_1988}%
  \BibitemOpen
  \bibfield  {author} {\bibinfo {author} {\bibfnamefont {W.}~\bibnamefont
  {von~der Linden}}\ and\ \bibinfo {author} {\bibfnamefont {P.}~\bibnamefont
  {Horsch}},\ }\href {\doibase 10.1103/PhysRevB.37.8351} {\bibfield  {journal}
  {\bibinfo  {journal} {Phys. Rev. B}\ }\textbf {\bibinfo {volume} {37}},\
  \bibinfo {pages} {8351} (\bibinfo {year} {1988})}\BibitemShut {NoStop}%
\bibitem [{\citenamefont {Northrup}, \citenamefont {Hybertsen},\ and\
  \citenamefont {Louie}(1991)}]{Northrup_1991}%
  \BibitemOpen
  \bibfield  {author} {\bibinfo {author} {\bibfnamefont {J.~E.}\ \bibnamefont
  {Northrup}}, \bibinfo {author} {\bibfnamefont {M.~S.}\ \bibnamefont
  {Hybertsen}}, \ and\ \bibinfo {author} {\bibfnamefont {S.~G.}\ \bibnamefont
  {Louie}},\ }\href {\doibase 10.1103/PhysRevLett.66.500} {\bibfield  {journal}
  {\bibinfo  {journal} {Phys. Rev. Lett.}\ }\textbf {\bibinfo {volume} {66}},\
  \bibinfo {pages} {500} (\bibinfo {year} {1991})}\BibitemShut {NoStop}%
\bibitem [{\citenamefont {Blase}, \citenamefont {Zhu},\ and\ \citenamefont
  {Louie}(1994)}]{Blase_1994}%
  \BibitemOpen
  \bibfield  {author} {\bibinfo {author} {\bibfnamefont {X.}~\bibnamefont
  {Blase}}, \bibinfo {author} {\bibfnamefont {X.}~\bibnamefont {Zhu}}, \ and\
  \bibinfo {author} {\bibfnamefont {S.~G.}\ \bibnamefont {Louie}},\ }\href
  {\doibase 10.1103/PhysRevB.49.4973} {\bibfield  {journal} {\bibinfo
  {journal} {Phys. Rev. B}\ }\textbf {\bibinfo {volume} {49}},\ \bibinfo
  {pages} {4973} (\bibinfo {year} {1994})}\BibitemShut {NoStop}%
\bibitem [{\citenamefont {Rohlfing}, \citenamefont {Kr{\"u}ger},\ and\
  \citenamefont {Pollmann}(1995)}]{Rohlfing_1995}%
  \BibitemOpen
  \bibfield  {author} {\bibinfo {author} {\bibfnamefont {M.}~\bibnamefont
  {Rohlfing}}, \bibinfo {author} {\bibfnamefont {P.}~\bibnamefont
  {Kr{\"u}ger}}, \ and\ \bibinfo {author} {\bibfnamefont {J.}~\bibnamefont
  {Pollmann}},\ }\href {\doibase 10.1103/PhysRevB.52.1905} {\bibfield
  {journal} {\bibinfo  {journal} {Phys. Rev. B}\ }\textbf {\bibinfo {volume}
  {52}},\ \bibinfo {pages} {1905} (\bibinfo {year} {1995})}\BibitemShut
  {NoStop}%
\bibitem [{\citenamefont {Peng}\ \emph {et~al.}(2013)\citenamefont {Peng},
  \citenamefont {Steinmann}, \citenamefont {{van Aggelen}},\ and\ \citenamefont
  {Yang}}]{Peng_2013}%
  \BibitemOpen
  \bibfield  {author} {\bibinfo {author} {\bibfnamefont {D.}~\bibnamefont
  {Peng}}, \bibinfo {author} {\bibfnamefont {S.~N.}\ \bibnamefont {Steinmann}},
  \bibinfo {author} {\bibfnamefont {H.}~\bibnamefont {{van Aggelen}}}, \ and\
  \bibinfo {author} {\bibfnamefont {W.}~\bibnamefont {Yang}},\ }\href {\doibase
  10.1063/1.4820556} {\bibfield  {journal} {\bibinfo  {journal} {J. Chem.
  Phys.}\ }\textbf {\bibinfo {volume} {139}},\ \bibinfo {pages} {104112}
  (\bibinfo {year} {2013})}\BibitemShut {NoStop}%
\bibitem [{\citenamefont {Scuseria}, \citenamefont {Henderson},\ and\
  \citenamefont {Bulik}(2013)}]{Scuseria_2013}%
  \BibitemOpen
  \bibfield  {author} {\bibinfo {author} {\bibfnamefont {G.~E.}\ \bibnamefont
  {Scuseria}}, \bibinfo {author} {\bibfnamefont {T.~M.}\ \bibnamefont
  {Henderson}}, \ and\ \bibinfo {author} {\bibfnamefont {I.~W.}\ \bibnamefont
  {Bulik}},\ }\href {\doibase 10.1063/1.4820557} {\bibfield  {journal}
  {\bibinfo  {journal} {J. Chem. Phys.}\ }\textbf {\bibinfo {volume} {139}},\
  \bibinfo {pages} {104113} (\bibinfo {year} {2013})}\BibitemShut {NoStop}%
\bibitem [{\citenamefont {M\"uller}, \citenamefont {Bl\"ugel},\ and\
  \citenamefont {Friedrich}(2019)}]{Muller_2019}%
  \BibitemOpen
  \bibfield  {author} {\bibinfo {author} {\bibfnamefont {M.~C. T.~D.}\
  \bibnamefont {M\"uller}}, \bibinfo {author} {\bibfnamefont {S.}~\bibnamefont
  {Bl\"ugel}}, \ and\ \bibinfo {author} {\bibfnamefont {C.}~\bibnamefont
  {Friedrich}},\ }\href {\doibase 10.1103/PhysRevB.100.045130} {\bibfield
  {journal} {\bibinfo  {journal} {Phys. Rev. B}\ }\textbf {\bibinfo {volume}
  {100}},\ \bibinfo {pages} {045130} (\bibinfo {year} {2019})}\BibitemShut
  {NoStop}%
\bibitem [{\citenamefont {M{\l}y{\'{n}}czak}\ \emph {et~al.}(2019)\citenamefont
  {M{\l}y{\'{n}}czak}, \citenamefont {M\"{u}ller}, \citenamefont
  {Gospodari{\v{c}}}, \citenamefont {Heider}, \citenamefont {Aguilera},
  \citenamefont {Bihlmayer}, \citenamefont {Gehlmann}, \citenamefont {Jugovac},
  \citenamefont {Zamborlini}, \citenamefont {Tusche}, \citenamefont {Suga},
  \citenamefont {Feyer}, \citenamefont {Plucinski}, \citenamefont {Friedrich},
  \citenamefont {Bl\"{u}gel},\ and\ \citenamefont {Schneider}}]{Myczak_2019}%
  \BibitemOpen
  \bibfield  {author} {\bibinfo {author} {\bibfnamefont {E.}~\bibnamefont
  {M{\l}y{\'{n}}czak}}, \bibinfo {author} {\bibfnamefont {M.~C. T.~D.}\
  \bibnamefont {M\"{u}ller}}, \bibinfo {author} {\bibfnamefont
  {P.}~\bibnamefont {Gospodari{\v{c}}}}, \bibinfo {author} {\bibfnamefont
  {T.}~\bibnamefont {Heider}}, \bibinfo {author} {\bibfnamefont
  {I.}~\bibnamefont {Aguilera}}, \bibinfo {author} {\bibfnamefont
  {G.}~\bibnamefont {Bihlmayer}}, \bibinfo {author} {\bibfnamefont
  {M.}~\bibnamefont {Gehlmann}}, \bibinfo {author} {\bibfnamefont
  {M.}~\bibnamefont {Jugovac}}, \bibinfo {author} {\bibfnamefont
  {G.}~\bibnamefont {Zamborlini}}, \bibinfo {author} {\bibfnamefont
  {C.}~\bibnamefont {Tusche}}, \bibinfo {author} {\bibfnamefont
  {S.}~\bibnamefont {Suga}}, \bibinfo {author} {\bibfnamefont {V.}~\bibnamefont
  {Feyer}}, \bibinfo {author} {\bibfnamefont {L.}~\bibnamefont {Plucinski}},
  \bibinfo {author} {\bibfnamefont {C.}~\bibnamefont {Friedrich}}, \bibinfo
  {author} {\bibfnamefont {S.}~\bibnamefont {Bl\"{u}gel}}, \ and\ \bibinfo
  {author} {\bibfnamefont {C.~M.}\ \bibnamefont {Schneider}},\ }\href {\doibase
  10.1038/s41467-019-08445-1} {\bibfield  {journal} {\bibinfo  {journal} {Nat.
  Comm.}\ }\textbf {\bibinfo {volume} {10}},\ \bibinfo {pages} {505} (\bibinfo
  {year} {2019})}\BibitemShut {NoStop}%
\bibitem [{\citenamefont {Friedrich}(2019)}]{Friedrich_2019}%
  \BibitemOpen
  \bibfield  {author} {\bibinfo {author} {\bibfnamefont {C.}~\bibnamefont
  {Friedrich}},\ }\href {\doibase 10.1103/PhysRevB.100.075142} {\bibfield
  {journal} {\bibinfo  {journal} {Phys. Rev. B}\ }\textbf {\bibinfo {volume}
  {100}},\ \bibinfo {pages} {075142} (\bibinfo {year} {2019})}\BibitemShut
  {NoStop}%
\bibitem [{\citenamefont {Nabok}, \citenamefont {Bl\"{u}gel},\ and\
  \citenamefont {Friedrich}(2021)}]{Nabok_2021}%
  \BibitemOpen
  \bibfield  {author} {\bibinfo {author} {\bibfnamefont {D.}~\bibnamefont
  {Nabok}}, \bibinfo {author} {\bibfnamefont {S.}~\bibnamefont {Bl\"{u}gel}}, \
  and\ \bibinfo {author} {\bibfnamefont {C.}~\bibnamefont {Friedrich}},\ }\href
  {\doibase 10.1038/s41524-021-00649-8} {\bibfield  {journal} {\bibinfo
  {journal} {npj Computational Materials}\ }\textbf {\bibinfo {volume} {7}},\
  \bibinfo {pages} {178} (\bibinfo {year} {2021})}\BibitemShut {NoStop}%
\bibitem [{\citenamefont {{van Setten}}\ \emph {et~al.}(2015)\citenamefont
  {{van Setten}}, \citenamefont {Caruso}, \citenamefont {Sharifzadeh},
  \citenamefont {Ren}, \citenamefont {Scheffler}, \citenamefont {Liu},
  \citenamefont {Lischner}, \citenamefont {Lin}, \citenamefont {Deslippe},
  \citenamefont {Louie}, \citenamefont {Yang}, \citenamefont {Weigend},
  \citenamefont {Neaton}, \citenamefont {Evers},\ and\ \citenamefont
  {Rinke}}]{vanSetten_2015}%
  \BibitemOpen
  \bibfield  {author} {\bibinfo {author} {\bibfnamefont {M.~J.}\ \bibnamefont
  {{van Setten}}}, \bibinfo {author} {\bibfnamefont {F.}~\bibnamefont
  {Caruso}}, \bibinfo {author} {\bibfnamefont {S.}~\bibnamefont {Sharifzadeh}},
  \bibinfo {author} {\bibfnamefont {X.}~\bibnamefont {Ren}}, \bibinfo {author}
  {\bibfnamefont {M.}~\bibnamefont {Scheffler}}, \bibinfo {author}
  {\bibfnamefont {F.}~\bibnamefont {Liu}}, \bibinfo {author} {\bibfnamefont
  {J.}~\bibnamefont {Lischner}}, \bibinfo {author} {\bibfnamefont
  {L.}~\bibnamefont {Lin}}, \bibinfo {author} {\bibfnamefont {J.~R.}\
  \bibnamefont {Deslippe}}, \bibinfo {author} {\bibfnamefont {S.~G.}\
  \bibnamefont {Louie}}, \bibinfo {author} {\bibfnamefont {C.}~\bibnamefont
  {Yang}}, \bibinfo {author} {\bibfnamefont {F.}~\bibnamefont {Weigend}},
  \bibinfo {author} {\bibfnamefont {J.~B.}\ \bibnamefont {Neaton}}, \bibinfo
  {author} {\bibfnamefont {F.}~\bibnamefont {Evers}}, \ and\ \bibinfo {author}
  {\bibfnamefont {P.}~\bibnamefont {Rinke}},\ }\href {\doibase
  10.1021/acs.jctc.5b00453} {\bibfield  {journal} {\bibinfo  {journal} {J.
  Chem. Theory Comput.}\ }\textbf {\bibinfo {volume} {11}},\ \bibinfo {pages}
  {5665} (\bibinfo {year} {2015})}\BibitemShut {NoStop}%
\bibitem [{\citenamefont {Loos}\ \emph {et~al.}(2020)\citenamefont {Loos},
  \citenamefont {Pradines}, \citenamefont {Scemama}, \citenamefont {Giner},\
  and\ \citenamefont {Toulouse}}]{Loos_2020}%
  \BibitemOpen
  \bibfield  {author} {\bibinfo {author} {\bibfnamefont {P.-F.}\ \bibnamefont
  {Loos}}, \bibinfo {author} {\bibfnamefont {B.}~\bibnamefont {Pradines}},
  \bibinfo {author} {\bibfnamefont {A.}~\bibnamefont {Scemama}}, \bibinfo
  {author} {\bibfnamefont {E.}~\bibnamefont {Giner}}, \ and\ \bibinfo {author}
  {\bibfnamefont {J.}~\bibnamefont {Toulouse}},\ }\href {\doibase
  10.1021/acs.jctc.9b01067} {\bibfield  {journal} {\bibinfo  {journal} {J.
  Chem. Theory Comput.}\ }\textbf {\bibinfo {volume} {16}},\ \bibinfo {pages}
  {1018} (\bibinfo {year} {2020})}\BibitemShut {NoStop}%
\bibitem [{\citenamefont {Monino}\ and\ \citenamefont
  {Loos}(2023)}]{Monino_2023}%
  \BibitemOpen
  \bibfield  {author} {\bibinfo {author} {\bibfnamefont {E.}~\bibnamefont
  {Monino}}\ and\ \bibinfo {author} {\bibfnamefont {P.-F.}\ \bibnamefont
  {Loos}},\ }\href {\doibase 10.1063/5.0159853} {\bibfield  {journal} {\bibinfo
   {journal} {J. Chem. Phys.}\ }\textbf {\bibinfo {volume} {159}},\ \bibinfo
  {pages} {034105} (\bibinfo {year} {2023})}\BibitemShut {NoStop}%
\bibitem [{\citenamefont {Krause}, \citenamefont {Harding},\ and\ \citenamefont
  {Klopper}(2015)}]{Krause_2015}%
  \BibitemOpen
  \bibfield  {author} {\bibinfo {author} {\bibfnamefont {K.}~\bibnamefont
  {Krause}}, \bibinfo {author} {\bibfnamefont {M.~E.}\ \bibnamefont {Harding}},
  \ and\ \bibinfo {author} {\bibfnamefont {W.}~\bibnamefont {Klopper}},\ }\href
  {\doibase 10.1080/00268976.2015.1025113} {\bibfield  {journal} {\bibinfo
  {journal} {Mol. Phys.}\ }\textbf {\bibinfo {volume} {113}},\ \bibinfo {pages}
  {1952} (\bibinfo {year} {2015})}\BibitemShut {NoStop}%
\bibitem [{\citenamefont {Springer}, \citenamefont {Aryasetiawan},\ and\
  \citenamefont {Karlsson}(1998)}]{Springer_1998}%
  \BibitemOpen
  \bibfield  {author} {\bibinfo {author} {\bibfnamefont {M.}~\bibnamefont
  {Springer}}, \bibinfo {author} {\bibfnamefont {F.}~\bibnamefont
  {Aryasetiawan}}, \ and\ \bibinfo {author} {\bibfnamefont {K.}~\bibnamefont
  {Karlsson}},\ }\href {\doibase 10.1103/PhysRevLett.80.2389} {\bibfield
  {journal} {\bibinfo  {journal} {Phys. Rev. Lett.}\ }\textbf {\bibinfo
  {volume} {80}},\ \bibinfo {pages} {2389} (\bibinfo {year}
  {1998})}\BibitemShut {NoStop}%
\bibitem [{\citenamefont {Zhukov}, \citenamefont {Chulkov},\ and\ \citenamefont
  {Echenique}(2004)}]{Zhukov_2004}%
  \BibitemOpen
  \bibfield  {author} {\bibinfo {author} {\bibfnamefont {V.~P.}\ \bibnamefont
  {Zhukov}}, \bibinfo {author} {\bibfnamefont {E.~V.}\ \bibnamefont {Chulkov}},
  \ and\ \bibinfo {author} {\bibfnamefont {P.~M.}\ \bibnamefont {Echenique}},\
  }\href {\doibase 10.1103/PhysRevLett.93.096401} {\bibfield  {journal}
  {\bibinfo  {journal} {Phys. Rev. Lett.}\ }\textbf {\bibinfo {volume} {93}},\
  \bibinfo {pages} {096401} (\bibinfo {year} {2004})}\BibitemShut {NoStop}%
\bibitem [{\citenamefont {Zhukov}, \citenamefont {Chulkov},\ and\ \citenamefont
  {Echenique}(2005)}]{Zhukov_2005}%
  \BibitemOpen
  \bibfield  {author} {\bibinfo {author} {\bibfnamefont {V.~P.}\ \bibnamefont
  {Zhukov}}, \bibinfo {author} {\bibfnamefont {E.~V.}\ \bibnamefont {Chulkov}},
  \ and\ \bibinfo {author} {\bibfnamefont {P.~M.}\ \bibnamefont {Echenique}},\
  }\href {\doibase 10.1103/PhysRevB.72.155109} {\bibfield  {journal} {\bibinfo
  {journal} {Phys. Rev. B}\ }\textbf {\bibinfo {volume} {72}},\ \bibinfo
  {pages} {72.155109} (\bibinfo {year} {2005})}\BibitemShut {NoStop}%
\bibitem [{\citenamefont {Zhukov}, \citenamefont {Chulkov},\ and\ \citenamefont
  {Echenique}(2006)}]{Zhukov_2006}%
  \BibitemOpen
  \bibfield  {author} {\bibinfo {author} {\bibfnamefont {V.~P.}\ \bibnamefont
  {Zhukov}}, \bibinfo {author} {\bibfnamefont {E.~V.}\ \bibnamefont {Chulkov}},
  \ and\ \bibinfo {author} {\bibfnamefont {P.~M.}\ \bibnamefont {Echenique}},\
  }\href {\doibase 10.1103/PhysRevB.73.125105} {\bibfield  {journal} {\bibinfo
  {journal} {Phys. Rev. B}\ }\textbf {\bibinfo {volume} {73}},\ \bibinfo
  {pages} {125105} (\bibinfo {year} {2006})}\BibitemShut {NoStop}%
\bibitem [{\citenamefont {Nechaev}\ and\ \citenamefont
  {Chulkov}(2006)}]{Nechaev_2006}%
  \BibitemOpen
  \bibfield  {author} {\bibinfo {author} {\bibfnamefont {I.~A.}\ \bibnamefont
  {Nechaev}}\ and\ \bibinfo {author} {\bibfnamefont {E.~V.}\ \bibnamefont
  {Chulkov}},\ }\href {\doibase 10.1103/PhysRevB.73.165112} {\bibfield
  {journal} {\bibinfo  {journal} {Phys. Rev. B}\ }\textbf {\bibinfo {volume}
  {73}},\ \bibinfo {pages} {165112} (\bibinfo {year} {2006})}\BibitemShut
  {NoStop}%
\bibitem [{\citenamefont {Nechaev}\ \emph {et~al.}(2008)\citenamefont
  {Nechaev}, \citenamefont {Sklyadneva}, \citenamefont {Silkin}, \citenamefont
  {Echenique},\ and\ \citenamefont {Chulkov}}]{Nechaev_2008}%
  \BibitemOpen
  \bibfield  {author} {\bibinfo {author} {\bibfnamefont {I.~A.}\ \bibnamefont
  {Nechaev}}, \bibinfo {author} {\bibfnamefont {I.~Y.}\ \bibnamefont
  {Sklyadneva}}, \bibinfo {author} {\bibfnamefont {V.~M.}\ \bibnamefont
  {Silkin}}, \bibinfo {author} {\bibfnamefont {P.~M.}\ \bibnamefont
  {Echenique}}, \ and\ \bibinfo {author} {\bibfnamefont {E.~V.}\ \bibnamefont
  {Chulkov}},\ }\href {\doibase 10.1103/PhysRevB.78.085113} {\bibfield
  {journal} {\bibinfo  {journal} {Phys. Rev. B}\ }\textbf {\bibinfo {volume}
  {78}},\ \bibinfo {pages} {085113} (\bibinfo {year} {2008})}\BibitemShut
  {NoStop}%
\bibitem [{\citenamefont {M\"onnich}\ \emph {et~al.}(2006)\citenamefont
  {M\"onnich}, \citenamefont {Lange}, \citenamefont {Bauer}, \citenamefont
  {Aeschlimann}, \citenamefont {Nechaev}, \citenamefont {Zhukov}, \citenamefont
  {Echenique},\ and\ \citenamefont {Chulkov}}]{Monnich_2006}%
  \BibitemOpen
  \bibfield  {author} {\bibinfo {author} {\bibfnamefont {A.}~\bibnamefont
  {M\"onnich}}, \bibinfo {author} {\bibfnamefont {J.}~\bibnamefont {Lange}},
  \bibinfo {author} {\bibfnamefont {M.}~\bibnamefont {Bauer}}, \bibinfo
  {author} {\bibfnamefont {M.}~\bibnamefont {Aeschlimann}}, \bibinfo {author}
  {\bibfnamefont {I.~A.}\ \bibnamefont {Nechaev}}, \bibinfo {author}
  {\bibfnamefont {V.~P.}\ \bibnamefont {Zhukov}}, \bibinfo {author}
  {\bibfnamefont {P.~M.}\ \bibnamefont {Echenique}}, \ and\ \bibinfo {author}
  {\bibfnamefont {E.~V.}\ \bibnamefont {Chulkov}},\ }\href {\doibase
  10.1103/PhysRevB.74.035102} {\bibfield  {journal} {\bibinfo  {journal} {Phys.
  Rev. B}\ }\textbf {\bibinfo {volume} {74}},\ \bibinfo {pages} {035102}
  (\bibinfo {year} {2006})}\BibitemShut {NoStop}%
\bibitem [{\citenamefont {Dreuw}\ and\ \citenamefont
  {Head-Gordon}(2005)}]{Dreuw_2005}%
  \BibitemOpen
  \bibfield  {author} {\bibinfo {author} {\bibfnamefont {A.}~\bibnamefont
  {Dreuw}}\ and\ \bibinfo {author} {\bibfnamefont {M.}~\bibnamefont
  {Head-Gordon}},\ }\href {\doibase 10.1021/cr0505627} {\bibfield  {journal}
  {\bibinfo  {journal} {Chem. Rev.}\ }\textbf {\bibinfo {volume} {105}},\
  \bibinfo {pages} {4009} (\bibinfo {year} {2005})}\BibitemShut {NoStop}%
\bibitem [{\citenamefont {Bauschlicher}\ and\ \citenamefont
  {Langhoff}(1987)}]{Bauschlicher_1987}%
  \BibitemOpen
  \bibfield  {author} {\bibinfo {author} {\bibfnamefont {J.}~\bibnamefont
  {Bauschlicher}, \bibfnamefont {Charles~W.}}\ and\ \bibinfo {author}
  {\bibfnamefont {S.~R.}\ \bibnamefont {Langhoff}},\ }\href {\doibase
  10.1063/1.453080} {\bibfield  {journal} {\bibinfo  {journal} {J. Chem.
  Phys.}\ }\textbf {\bibinfo {volume} {87}},\ \bibinfo {pages} {2919} (\bibinfo
  {year} {1987})}\BibitemShut {NoStop}%
\bibitem [{\citenamefont {Bruna}\ and\ \citenamefont
  {Wright}(1990)}]{Bruna_1990}%
  \BibitemOpen
  \bibfield  {author} {\bibinfo {author} {\bibfnamefont {P.~J.}\ \bibnamefont
  {Bruna}}\ and\ \bibinfo {author} {\bibfnamefont {J.~S.}\ \bibnamefont
  {Wright}},\ }\href {\doibase 10.1021/j100368a014} {\bibfield  {journal}
  {\bibinfo  {journal} {J. Phys. Chem.}\ }\textbf {\bibinfo {volume} {94}},\
  \bibinfo {pages} {1774} (\bibinfo {year} {1990})}\BibitemShut {NoStop}%
\bibitem [{\citenamefont {Abrams}\ and\ \citenamefont
  {Sherrill}(2004)}]{Abrams_2004}%
  \BibitemOpen
  \bibfield  {author} {\bibinfo {author} {\bibfnamefont {M.~L.}\ \bibnamefont
  {Abrams}}\ and\ \bibinfo {author} {\bibfnamefont {C.~D.}\ \bibnamefont
  {Sherrill}},\ }\href {\doibase 10.1063/1.1804498} {\bibfield  {journal}
  {\bibinfo  {journal} {J. Chem. Phys.}\ }\textbf {\bibinfo {volume} {121}},\
  \bibinfo {pages} {9211} (\bibinfo {year} {2004})}\BibitemShut {NoStop}%
\bibitem [{\citenamefont {Sherrill}\ and\ \citenamefont
  {Piecuch}(2005)}]{Sherrill_2005}%
  \BibitemOpen
  \bibfield  {author} {\bibinfo {author} {\bibfnamefont {C.~D.}\ \bibnamefont
  {Sherrill}}\ and\ \bibinfo {author} {\bibfnamefont {P.}~\bibnamefont
  {Piecuch}},\ }\href {\doibase 10.1063/1.1867379} {\bibfield  {journal}
  {\bibinfo  {journal} {J. Chem. Phys.}\ }\textbf {\bibinfo {volume} {122}},\
  \bibinfo {pages} {124104} (\bibinfo {year} {2005})}\BibitemShut {NoStop}%
\bibitem [{\citenamefont {Li}\ and\ \citenamefont {Paldus}(2006)}]{Li_2006}%
  \BibitemOpen
  \bibfield  {author} {\bibinfo {author} {\bibfnamefont {X.}~\bibnamefont
  {Li}}\ and\ \bibinfo {author} {\bibfnamefont {J.}~\bibnamefont {Paldus}},\
  }\href {\doibase
  https://doi-org-s.docadis.univ-tlse3.fr/10.1016/j.cplett.2006.09.053}
  {\bibfield  {journal} {\bibinfo  {journal} {Chem. Phys. Lett.}\ }\textbf
  {\bibinfo {volume} {431}},\ \bibinfo {pages} {179} (\bibinfo {year}
  {2006})}\BibitemShut {NoStop}%
\bibitem [{\citenamefont {Booth}\ \emph {et~al.}(2011)\citenamefont {Booth},
  \citenamefont {Cleland}, \citenamefont {Thom},\ and\ \citenamefont
  {Alavi}}]{Booth_2011}%
  \BibitemOpen
  \bibfield  {author} {\bibinfo {author} {\bibfnamefont {G.~H.}\ \bibnamefont
  {Booth}}, \bibinfo {author} {\bibfnamefont {D.}~\bibnamefont {Cleland}},
  \bibinfo {author} {\bibfnamefont {A.~J.~W.}\ \bibnamefont {Thom}}, \ and\
  \bibinfo {author} {\bibfnamefont {A.}~\bibnamefont {Alavi}},\ }\href
  {\doibase 10.1063/1.3624383} {\bibfield  {journal} {\bibinfo  {journal} {J.
  Chem. Phys.}\ }\textbf {\bibinfo {volume} {135}},\ \bibinfo {pages} {084104}
  (\bibinfo {year} {2011})}\BibitemShut {NoStop}%
\bibitem [{\citenamefont {Huber}\ and\ \citenamefont
  {Herzberg}(1979)}]{HerzbergBook}%
  \BibitemOpen
  \bibfield  {author} {\bibinfo {author} {\bibfnamefont {K.~P.}\ \bibnamefont
  {Huber}}\ and\ \bibinfo {author} {\bibfnamefont {G.}~\bibnamefont
  {Herzberg}},\ }\href@noop {} {\emph {\bibinfo {title} {Molecular Spectra and
  Molecular Structure: IV. Constants of diatomic molecules}}}\ (\bibinfo
  {publisher} {van Nostrand Reinhold Company},\ \bibinfo {year}
  {1979})\BibitemShut {NoStop}%
\bibitem [{\citenamefont {Garniron}\ \emph {et~al.}(2019)\citenamefont
  {Garniron}, \citenamefont {Gasperich}, \citenamefont {Applencourt},
  \citenamefont {Benali}, \citenamefont {Fert{\'e}}, \citenamefont {Paquier},
  \citenamefont {Pradines}, \citenamefont {Assaraf}, \citenamefont {Reinhardt},
  \citenamefont {Toulouse}, \citenamefont {Barbaresco}, \citenamefont {Renon},
  \citenamefont {David}, \citenamefont {Malrieu}, \citenamefont {V{\'e}ril},
  \citenamefont {Caffarel}, \citenamefont {Loos}, \citenamefont {Giner},\ and\
  \citenamefont {Scemama}}]{Garniron_2019}%
  \BibitemOpen
  \bibfield  {author} {\bibinfo {author} {\bibfnamefont {Y.}~\bibnamefont
  {Garniron}}, \bibinfo {author} {\bibfnamefont {K.}~\bibnamefont {Gasperich}},
  \bibinfo {author} {\bibfnamefont {T.}~\bibnamefont {Applencourt}}, \bibinfo
  {author} {\bibfnamefont {A.}~\bibnamefont {Benali}}, \bibinfo {author}
  {\bibfnamefont {A.}~\bibnamefont {Fert{\'e}}}, \bibinfo {author}
  {\bibfnamefont {J.}~\bibnamefont {Paquier}}, \bibinfo {author} {\bibfnamefont
  {B.}~\bibnamefont {Pradines}}, \bibinfo {author} {\bibfnamefont
  {R.}~\bibnamefont {Assaraf}}, \bibinfo {author} {\bibfnamefont
  {P.}~\bibnamefont {Reinhardt}}, \bibinfo {author} {\bibfnamefont
  {J.}~\bibnamefont {Toulouse}}, \bibinfo {author} {\bibfnamefont
  {P.}~\bibnamefont {Barbaresco}}, \bibinfo {author} {\bibfnamefont
  {N.}~\bibnamefont {Renon}}, \bibinfo {author} {\bibfnamefont
  {G.}~\bibnamefont {David}}, \bibinfo {author} {\bibfnamefont {J.~P.}\
  \bibnamefont {Malrieu}}, \bibinfo {author} {\bibfnamefont {M.}~\bibnamefont
  {V{\'e}ril}}, \bibinfo {author} {\bibfnamefont {M.}~\bibnamefont {Caffarel}},
  \bibinfo {author} {\bibfnamefont {P.~F.}\ \bibnamefont {Loos}}, \bibinfo
  {author} {\bibfnamefont {E.}~\bibnamefont {Giner}}, \ and\ \bibinfo {author}
  {\bibfnamefont {A.}~\bibnamefont {Scemama}},\ }\href {\doibase
  10.1021/acs.jctc.9b00176} {\bibfield  {journal} {\bibinfo  {journal} {J.
  Chem. Theory Comput.}\ }\textbf {\bibinfo {volume} {15}},\ \bibinfo {pages}
  {3591} (\bibinfo {year} {2019})}\BibitemShut {NoStop}%
\bibitem [{\citenamefont {Bickers}, \citenamefont {Scalapino},\ and\
  \citenamefont {White}(1989)}]{Bickers_1989a}%
  \BibitemOpen
  \bibfield  {author} {\bibinfo {author} {\bibfnamefont {N.~E.}\ \bibnamefont
  {Bickers}}, \bibinfo {author} {\bibfnamefont {D.~J.}\ \bibnamefont
  {Scalapino}}, \ and\ \bibinfo {author} {\bibfnamefont {S.~R.}\ \bibnamefont
  {White}},\ }\href {\doibase 10.1103/PhysRevLett.62.961} {\bibfield  {journal}
  {\bibinfo  {journal} {Phys. Rev. Lett.}\ }\textbf {\bibinfo {volume} {62}},\
  \bibinfo {pages} {961} (\bibinfo {year} {1989})}\BibitemShut {NoStop}%
\bibitem [{\citenamefont {Bickers}\ and\ \citenamefont
  {Scalapino}(1989)}]{Bickers_1989b}%
  \BibitemOpen
  \bibfield  {author} {\bibinfo {author} {\bibfnamefont {N.}~\bibnamefont
  {Bickers}}\ and\ \bibinfo {author} {\bibfnamefont {D.}~\bibnamefont
  {Scalapino}},\ }\href {\doibase https://doi.org/10.1016/0003-4916(89)90359-X}
  {\bibfield  {journal} {\bibinfo  {journal} {Ann. Phys.}\ }\textbf {\bibinfo
  {volume} {193}},\ \bibinfo {pages} {206} (\bibinfo {year}
  {1989})}\BibitemShut {NoStop}%
\bibitem [{\citenamefont {Bickers}\ and\ \citenamefont
  {White}(1991)}]{Bickers_1991}%
  \BibitemOpen
  \bibfield  {author} {\bibinfo {author} {\bibfnamefont {N.~E.}\ \bibnamefont
  {Bickers}}\ and\ \bibinfo {author} {\bibfnamefont {S.~R.}\ \bibnamefont
  {White}},\ }\href {\doibase 10.1103/PhysRevB.43.8044} {\bibfield  {journal}
  {\bibinfo  {journal} {Phys. Rev. B}\ }\textbf {\bibinfo {volume} {43}},\
  \bibinfo {pages} {8044} (\bibinfo {year} {1991})}\BibitemShut {NoStop}%
\bibitem [{\citenamefont {Bickers}(2004)}]{Bickers_2004}%
  \BibitemOpen
  \bibfield  {author} {\bibinfo {author} {\bibfnamefont {N.~E.}\ \bibnamefont
  {Bickers}},\ }\enquote {\bibinfo {title} {Self-consistent many-body theory
  for condensed matter systems},}\ in\ \href {\doibase 10.1007/0-387-21717-7_6}
  {\emph {\bibinfo {booktitle} {Theoretical Methods for Strongly Correlated
  Electrons}}},\ \bibinfo {editor} {edited by\ \bibinfo {editor} {\bibfnamefont
  {D.}~\bibnamefont {S{\'e}n{\'e}chal}}, \bibinfo {editor} {\bibfnamefont
  {A.-M.}\ \bibnamefont {Tremblay}}, \ and\ \bibinfo {editor} {\bibfnamefont
  {C.}~\bibnamefont {Bourbonnais}}}\ (\bibinfo  {publisher} {Springer New
  York},\ \bibinfo {year} {2004})\ pp.\ \bibinfo {pages} {237--296}\BibitemShut
  {NoStop}%
\bibitem [{\citenamefont {De~Dominicis}\ and\ \citenamefont
  {Martin}(1964{\natexlab{a}})}]{DeDominicis_1964a}%
  \BibitemOpen
  \bibfield  {author} {\bibinfo {author} {\bibfnamefont {C.}~\bibnamefont
  {De~Dominicis}}\ and\ \bibinfo {author} {\bibfnamefont {P.~C.}\ \bibnamefont
  {Martin}},\ }\href {\doibase 10.1063/1.1704062} {\bibfield  {journal}
  {\bibinfo  {journal} {J. Math. Phys.}\ }\textbf {\bibinfo {volume} {5}},\
  \bibinfo {pages} {14} (\bibinfo {year} {1964}{\natexlab{a}})}\BibitemShut
  {NoStop}%
\bibitem [{\citenamefont {De~Dominicis}\ and\ \citenamefont
  {Martin}(1964{\natexlab{b}})}]{DeDominicis_1964b}%
  \BibitemOpen
  \bibfield  {author} {\bibinfo {author} {\bibfnamefont {C.}~\bibnamefont
  {De~Dominicis}}\ and\ \bibinfo {author} {\bibfnamefont {P.~C.}\ \bibnamefont
  {Martin}},\ }\href {\doibase 10.1063/1.1704064} {\bibfield  {journal}
  {\bibinfo  {journal} {J. Math. Phys.}\ }\textbf {\bibinfo {volume} {5}},\
  \bibinfo {pages} {31} (\bibinfo {year} {1964}{\natexlab{b}})}\BibitemShut
  {NoStop}%
\bibitem [{\citenamefont {Shepherd}, \citenamefont {Henderson},\ and\
  \citenamefont {Scuseria}(2014)}]{Shepherd_2014}%
  \BibitemOpen
  \bibfield  {author} {\bibinfo {author} {\bibfnamefont {J.~J.}\ \bibnamefont
  {Shepherd}}, \bibinfo {author} {\bibfnamefont {T.~M.}\ \bibnamefont
  {Henderson}}, \ and\ \bibinfo {author} {\bibfnamefont {G.~E.}\ \bibnamefont
  {Scuseria}},\ }\href {\doibase 10.1103/PhysRevLett.112.133002} {\bibfield
  {journal} {\bibinfo  {journal} {Phys. Rev. Lett.}\ }\textbf {\bibinfo
  {volume} {112}},\ \bibinfo {pages} {133002} (\bibinfo {year}
  {2014})}\BibitemShut {NoStop}%
\bibitem [{\citenamefont {Hofierka}\ \emph {et~al.}(2022)\citenamefont
  {Hofierka}, \citenamefont {Cunningham}, \citenamefont {Rawlins},
  \citenamefont {Patterson},\ and\ \citenamefont {Green}}]{Hofierka_2022}%
  \BibitemOpen
  \bibfield  {author} {\bibinfo {author} {\bibfnamefont {J.}~\bibnamefont
  {Hofierka}}, \bibinfo {author} {\bibfnamefont {B.}~\bibnamefont
  {Cunningham}}, \bibinfo {author} {\bibfnamefont {C.~M.}\ \bibnamefont
  {Rawlins}}, \bibinfo {author} {\bibfnamefont {C.~H.}\ \bibnamefont
  {Patterson}}, \ and\ \bibinfo {author} {\bibfnamefont {D.~G.}\ \bibnamefont
  {Green}},\ }\href {\doibase 10.1038/s41586-022-04703-3} {\bibfield  {journal}
  {\bibinfo  {journal} {Nature}\ }\textbf {\bibinfo {volume} {606}},\ \bibinfo
  {pages} {688} (\bibinfo {year} {2022})}\BibitemShut {NoStop}%
\bibitem [{\citenamefont {Riva}\ \emph {et~al.}(2022)\citenamefont {Riva},
  \citenamefont {Audinet}, \citenamefont {Vladaj}, \citenamefont {Romaniello},\
  and\ \citenamefont {Berger}}]{Riva_2022}%
  \BibitemOpen
  \bibfield  {author} {\bibinfo {author} {\bibfnamefont {G.}~\bibnamefont
  {Riva}}, \bibinfo {author} {\bibfnamefont {T.}~\bibnamefont {Audinet}},
  \bibinfo {author} {\bibfnamefont {M.}~\bibnamefont {Vladaj}}, \bibinfo
  {author} {\bibfnamefont {P.}~\bibnamefont {Romaniello}}, \ and\ \bibinfo
  {author} {\bibfnamefont {J.~A.}\ \bibnamefont {Berger}},\ }\href {\doibase
  10.21468/SciPostPhys.12.3.093} {\bibfield  {journal} {\bibinfo  {journal}
  {SciPost Phys.}\ }\textbf {\bibinfo {volume} {12}},\ \bibinfo {pages} {093}
  (\bibinfo {year} {2022})}\BibitemShut {NoStop}%
\bibitem [{\citenamefont {Riva}, \citenamefont {Romaniello},\ and\
  \citenamefont {Berger}(2023)}]{Riva_2023}%
  \BibitemOpen
  \bibfield  {author} {\bibinfo {author} {\bibfnamefont {G.}~\bibnamefont
  {Riva}}, \bibinfo {author} {\bibfnamefont {P.}~\bibnamefont {Romaniello}}, \
  and\ \bibinfo {author} {\bibfnamefont {J.~A.}\ \bibnamefont {Berger}},\
  }\href@noop {} {\enquote {\bibinfo {title} {{The multi-channel Dyson
  equation: coupling many-body Green's functions}},}\ }\bibinfo {howpublished}
  {under review} (\bibinfo {year} {2023})\BibitemShut {NoStop}%
\bibitem [{\citenamefont {Martin}\ and\ \citenamefont
  {Schwinger}(1959)}]{Martin_1959}%
  \BibitemOpen
  \bibfield  {author} {\bibinfo {author} {\bibfnamefont {P.~C.}\ \bibnamefont
  {Martin}}\ and\ \bibinfo {author} {\bibfnamefont {J.}~\bibnamefont
  {Schwinger}},\ }\href {\doibase 10.1103/PhysRev.115.1342} {\bibfield
  {journal} {\bibinfo  {journal} {Phys. Rev.}\ }\textbf {\bibinfo {volume}
  {115}},\ \bibinfo {pages} {1342} (\bibinfo {year} {1959})}\BibitemShut
  {NoStop}%
\bibitem [{\citenamefont {Angyan}\ \emph {et~al.}(2011)\citenamefont {Angyan},
  \citenamefont {Liu}, \citenamefont {Toulouse},\ and\ \citenamefont
  {Jansen}}]{Angyan_2011}%
  \BibitemOpen
  \bibfield  {author} {\bibinfo {author} {\bibfnamefont {J.~G.}\ \bibnamefont
  {Angyan}}, \bibinfo {author} {\bibfnamefont {R.-F.}\ \bibnamefont {Liu}},
  \bibinfo {author} {\bibfnamefont {J.}~\bibnamefont {Toulouse}}, \ and\
  \bibinfo {author} {\bibfnamefont {G.}~\bibnamefont {Jansen}},\ }\href
  {\doibase 10.1021/ct200501r} {\bibfield  {journal} {\bibinfo  {journal} {J.
  Chem. Theory Comput.}\ }\textbf {\bibinfo {volume} {7}},\ \bibinfo {pages}
  {3116} (\bibinfo {year} {2011})}\BibitemShut {NoStop}%
\bibitem [{\citenamefont {Monino}\ and\ \citenamefont
  {Loos}(2021)}]{Monino_2021}%
  \BibitemOpen
  \bibfield  {author} {\bibinfo {author} {\bibfnamefont {E.}~\bibnamefont
  {Monino}}\ and\ \bibinfo {author} {\bibfnamefont {P.-F.}\ \bibnamefont
  {Loos}},\ }\href {\doibase 10.1021/acs.jctc.1c00074} {\bibfield  {journal}
  {\bibinfo  {journal} {J. Chem. Theory Comput.}\ }\textbf {\bibinfo {volume}
  {17}},\ \bibinfo {pages} {2852} (\bibinfo {year} {2021})}\BibitemShut
  {NoStop}%
\bibitem [{\citenamefont {Yang}\ \emph {et~al.}(2013)\citenamefont {Yang},
  \citenamefont {{van Aggelen}}, \citenamefont {Steinmann}, \citenamefont
  {Peng},\ and\ \citenamefont {Yang}}]{Yang_2013}%
  \BibitemOpen
  \bibfield  {author} {\bibinfo {author} {\bibfnamefont {Y.}~\bibnamefont
  {Yang}}, \bibinfo {author} {\bibfnamefont {H.}~\bibnamefont {{van Aggelen}}},
  \bibinfo {author} {\bibfnamefont {S.~N.}\ \bibnamefont {Steinmann}}, \bibinfo
  {author} {\bibfnamefont {D.}~\bibnamefont {Peng}}, \ and\ \bibinfo {author}
  {\bibfnamefont {W.}~\bibnamefont {Yang}},\ }\href {\doibase
  10.1063/1.4828728} {\bibfield  {journal} {\bibinfo  {journal} {J. Chem.
  Phys.}\ }\textbf {\bibinfo {volume} {139}},\ \bibinfo {pages} {174110}
  (\bibinfo {year} {2013})}\BibitemShut {NoStop}%
\end{thebibliography}%

\end{document}